\documentclass[twocolumn, trackchanges]{aastex631}
\usepackage{array}

\graphicspath{{figure/}}

\usepackage{graphicx}
\usepackage{amsmath}
\usepackage{amssymb}	
\usepackage{float}
\usepackage{appendix}
\usepackage{macros}
\usepackage{chngcntr}
\usepackage{multirow}
\usepackage{newtxtext,newtxmath}
\usepackage[T1]{fontenc}
\hypersetup{linkcolor=red,citecolor=blue,filecolor=cyan,urlcolor=magenta}
\usepackage{xcolor}

\newcommand{\ucmerced}{Department of Physics, University of California, Merced, CA 95343, USA}
\newcommand{\ucdavis}{Department of Physics \& Astronomy, University of California, Davis, CA 95616, USA}

\begin{document}

\title[Abundances of Open Clusters in FIRE]{Strong Chemical Tagging in FIRE: Intra and Inter-Cluster Chemical Homogeneity in Open Clusters in Milky Way-like Galaxy Simulations}

\shorttitle{Intra and Inter-Cluster Chemical Homogeneity in Open Clusters}
\shortauthors{Bhattarai et al.}

\correspondingauthor{Binod Bhattarai}
\email{bbhattarai@ucmerced.edu}

\author[0000-0002-7707-1996]{Binod Bhattarai}
\affiliation{\ucmerced}

\author[0000-0003-3217-5967]{Sarah R.~Loebman}
\affiliation{\ucmerced}

\author[0000-0001-5082-6693]{Melissa K.~Ness}
\affiliation{Research School of Astronomy $\&$ Astrophysics, Australian National University, Canberra, ACT 2611, Australia}
\affiliation{Department of Astronomy, Columbia University, 550 West 120th Street, New York, NY, 10027, USA}

\author[0000-0003-0603-8942]{Andrew Wetzel}
\affiliation{\ucdavis}

\author[0000-0002-6993-0826]{Emily C. Cunningham}
\altaffiliation{NASA Hubble Fellow}
\affiliation{Department of Astronomy, Columbia University, 550 West 120th Street, New York, NY, 10027, USA}

\author[0009-0007-3431-4269]{Hanna Parul}
\affiliation{Department of Physics \& Astronomy, University of Alabama, Box 870324, Tuscaloosa, AL 35487-0324, USA}

\author[0009-0008-0081-764X]{Alessa Ibrahim Wiggins}
\affiliation{Department of Physics and Astronomy, Texas Christian University, Fort Worth, TX}

\begin{abstract}
Open star clusters are the essential building blocks of the Galactic disk; “strong chemical tagging” – the premise that all star clusters can be reconstructed given chemistry information alone – is a driving force behind many current and upcoming large Galactic spectroscopic surveys. 
In this work, we characterize abundance patterns for 9 elements (C, N, O, Ne, Mg, Si, S, Ca, and Fe) 
in open clusters (OCs) in three galaxies (m12i, m12f, and m12m) from the \textit{Latte} suite of FIRE-2 simulations to investigate if strong chemical tagging is possible in these simulations. 
We select young massive ($\ge$$10^{4.6} M_{\odot}$) OCs formed in the last $\sim$100 Myr and calculate the intra- and inter-cluster abundance scatter for these clusters.
We compare these results with analogous calculations drawn from observations of OCs in the Milky Way. 
We find the intra-cluster scatter of the observations and simulations to be comparable. 
While the abundance scatter within
each cluster is minimal ($\lesssim$$0.020$ dex), the mean abundance patterns of different clusters are not unique. 
We also calculate the chemical difference in intra- and inter-cluster star pairs and find it, in general, to be so small that it is difficult to distinguish between stars drawn from the same OC or from different OCs. 
Despite tracing three distinct nucleosynthetic families (core-collapse supernovae, white dwarf supernovae, and stellar winds), we conclude that 
these elemental abundances do not provide enough discriminating information to use strong chemical tagging for reliable OC membership.
\end{abstract}


\section{Introduction} \label{sec:introduction} 

Uncovering the history of the Milky Way's stellar disk is a fundamental goal of Galactic astronomy and is tied to the smallest scale structures where star formation occurs, open clusters (OCs).
OCs are groups of stars that are born together and remain gravitationally bound for a period of time;
\citet{Lada2003} define an OC as a group of young stars that survive tidal disruption for a period as long as 100 Myr. 
In numerical simulations, OCs are usually defined as groups of co-forming stars that share a common origin \citep[e.g.,][]{dobbs_2017,grudic_2023_great_balls_FIRE}.

OCs are often leveraged to study the dynamical and chemical history of the Galactic disk, as these bright structures are relatively straightforward to observe and characterize \citep[e.g.,][and references therein]{frinchaboy_2008,natalie_myers_2022_OCCAM}. 
OCs are considered the building blocks of the Galactic disk because a significant fraction of star formation occurs within them and the majority are thought to dissolve quickly \citep{Lada2003,bland_hawthorn_2010_evolution_of_the_galactic_disk,star_clusters_kruijssen_simulations,star_clusters_krumholz_2019}.
One of the important objectives of near-field astronomy is to reconstruct the structures that have dispersed due to dynamical processes in the disk, specifically, reassembling structures like OCs using present-day stellar observables.

One stellar observable of particular interest is elemental abundance.
 This is because stars that belong to a particular OC are thought to form from a common well-mixed molecular cloud with a similar underlying chemistry \citep{shu_1987_star_formation_molecular_clouds,meyer_2000_stellar_init_mass_functions_clusters,portegies_young)massive_star_clusters_2010,feng_krumholz_2014_early_turbulant_mixing_homogeneity_OC_simulations}. 
Based on this assumption, stars that were born within a given OC are believed to have a similar chemical ``fingerprint'' that would enable OC reconstruction based on chemistry alone \citep[e.g.,][]{Freeman_Hawthorn_2002_Metallicity,d_silva_2007_chemical_homogeneity,de_silva_abundance_patterns_2009,bland_hawthorn_2010_evolution_of_the_galactic_disk,jo_bovy_2016_chemical_homogeneity_of_open_clusters,manea_chemical_homogeneity}. This is the premise behind strong chemical tagging \citep{Freeman_Hawthorn_2002_Metallicity}, which relies on accurate measurement of multiple elements in stars to reconstruct cluster membership \citep[e.g.,][]{majewski_2012_chemistry,martell_2016_chemical_tagging,hogg_2016_chemical_tagging,jo_bovy_2016_chemical_homogeneity_of_open_clusters,spina_2022_chemical_tagging}.

The feasibility of strong chemical tagging is a key motivation for several current and upcoming spectroscopic surveys \citep[e.g., GALAH, APOGEE, WEAVE][]{GALAH,Buder_2019_GALAH,Majewski_APOGEE_2017, WEAVE, SDSS-V}.
For strong chemical tagging to work, the member stars within an OC should be nearly chemically homogeneous, i.e., they should have \emph{small intra-cluster dispersion}. 
In addition, each cluster should have a unique chemical signature so that it can be well distinguished from the stars that belong to other OCs, i.e., there should be \emph{significant inter-cluster dispersion} \citep{lambert_reddy_2016_inter_cluster_dispersion}.

Whether there is typically a small level of intra-cluster dispersion (i.e., the level of similarity in the element abundances of member stars) in OCs is still a matter of debate \citep{blanco_cuaresma_2015_inhomogeneity_evolutionary_stages,price_jones_bovy_2019_chemical_tagging}.
However, in the limit of small samples, the chemistry of stars that belong to a particular OC has often been shown to share a similar signature when compared with the chemistry of stars that belong to different clusters \citep[e.g.,][]{mitschang_2013_distance_metric,majewski_2012_chemistry,martell_2016_chemical_tagging,hogg_2016_chemical_tagging, de_silva_2006_chemical_homogeneity_hyades, pancino_2010_3_red_clump_stars_5_OCs_homogeneity}. 
This indicates that stars that belong to an OC could have a measurable signature that is distinct from other OCs, in the limit that measurement uncertainties in stellar abundances are smaller than the level of the intrinsic element abundance scatter, so that the amplitude of homogeneity in OCs can be uniquely assessed \citep{ness_2015_the_cannon_data_driven_approach_to_stellar_label_determination,ness_2018_doppelgangers}.

Large surveys have extended studies of OCs to more substantial samples and determined the intrinsic scatter to be $<$ 0.03 dex within clusters for up to 15 element abundances \citep{jo_bovy_2016_chemical_homogeneity_of_open_clusters, ness_2018_doppelgangers, bertran_de_lis_2016_variance_ofe, poovelil_2020_oc_chemical_homogeneity_MW}.  
Interestingly, \citet{blanco_cuaresma_2015_inhomogeneity_evolutionary_stages} find distinct chemical signatures for stars in different evolutionary stages that belong to the same open cluster. 
In fact, theoretical expectations lead to a prediction that intra-cluster scatter is non-zero when processes like non-thermal equilibrium, atomic diffusion, mixing, the influence of binarity and planetary engulfment are included in stellar abundance calculations.
These predictions have been born out both observationally and in theoretical models \citep{liu_2019_atomic_diffusion,dotter_2017}.

Besides debates about intra-cluster homogeneity, there are also several discussions in the literature about the level of inter-cluster homogeneity throughout the Milky Way. 
For example, there is an expectation that the inter-cluster homogeneity could be a function of galactocentric radius \citep{spina_2022_challenges_OCs}. 
In fact, a negative radial gradient in the Milky Way's disk has been found in observations \citep[e.g.,][]{boeche_2013_chemical_gradients_in_MW,anders_2014_chemodynamics_MW_APOGEE,donor_2020_OC_chemical_abundances,natalie_myers_2022_OCCAM,magrini_2023_abundance_gradient_OCs}, while radial and  azimuthal variations have been detected in several nearby galaxies as well \citep[e.g.,][]{sanchez_menguiano_2016_shape_of_oxygen_abundance_in_spiral_galaxies,molla_2019_2d_chemical_evolution_MW_type_galaxies,kreckel_2020_scale_of_ISM_in_spiral_galaxies,li_zefeng_2023_metallicity_distribution_galaxies}, which could be taken as an indication for the uniqueness of abundances in star clusters found at particular galactocentric location. 
However, \citet{blanco_cuaresma_2015_inhomogeneity_evolutionary_stages} find significant overlap of multiple chemical abundances in their sample of 31 OCs within the Milky Way.  
Moreover, \citet{ting_conroy_2015_inter_cluster} show that overdensities in chemical space do not guarantee that such overdensities arise due to a single set of stars from a common birth cloud, thus suggesting an overlap in the inter-cluster metallicity distribution. 

A differing perspective has been found by \citet{lambert_reddy_2016_inter_cluster_dispersion}, who show significant inter-cluster variation in the chemical composition of heavy elements (La, Ce, Nd, and Sm) for red giants in a sample of 28 OCs, where all the clusters had nearly solar metallicity. 
Moreover, \citet{manea_2023_distinguishing_power_neutron_capture} find that neutron-capture elements carry a more discerning signature than the lighter elements that have been traditionally considered in the current generation of spectroscopic surveys.
However, in the limit of elemental abundances available in APOGEE \citep{Majewski2016}, \citet{ness_2018_doppelgangers} find field stars have similar chemistry as the members of OCs, suggesting an overlap of birth cluster signatures with the field stars. Furthermore, in a different study, \citet{Ness_2019_feh_age} find that, at a fixed [Fe/H], chemistry is a
deterministic property of age that does not necessarily change with the birth location.

The perhaps most critical consideration in assessing the viability of chemical tagging is not simply how chemically homogeneous OCs are themselves, but how homogeneous stars within a cluster are relative to random field stars. 
The contamination rate using 20 abundances in \citet{ness_2018_doppelgangers} was found to be around 1 percent at [Fe/H] = 0. 
That is, while the intra-cluster dispersion is small, on the order of $<$ 0.03 dex, around 1 percent of random field stars at solar [Fe/H] are as chemically similar as stars within the same cluster.
This is very prohibitive for any chemical tagging pursuit. 
However, what remains critically unclear is the theoretical expectation for inter and intra-cluster homogeneity from simulations. 
In this work, we test this expectation using galaxy simulations that resolve OCs in a cosmological context. 

Previous simulation studies have shown that turbulent mixing plays an important role homogenizing molecular clouds and reducing stellar abundance scatter \citep{McKee_Tan_2002_GMC_well_mixed}.
This can be particularly significant during cloud assembly, potentially explaining the observed chemical homogeneity of stars that originate from the same molecular cloud \citep{feng_krumholz_2014_early_turbulant_mixing_homogeneity_OC_simulations}. 
Extending this work, \citet{armillotta_krumholz_2018_mixing_of_metals_simulations} zoom in on the collapse of one giant molecular cloud extracted from a galaxy simulation and study the formation of individual stars down to a spatial resolution of $\approx$$10^{-3}$ pc. 
They find that the star formation process defines a natural size scale of $\sim$1 pc for chemically homogeneous star clusters, suggesting stars within $\sim$1 pc of each other share similar chemical properties. 
However, these simulations assume metals are well coupled with gas, overlooking potential effects of metals in the form of dust grains. 
Taking a different approach to identify co-natal stars, \citet{kamdar_2019_dynamical_model_clustered_star_formation} evolve 4 billion stars over 5 Gyr in a realistic potential with various galactic structures. 
They find that combining chemical and phase space information enhances the identification of co-natal populations, suggesting co-moving pairs of stars with velocity separation $<$2 km/s and metallicity separation $<$0.05 dex are likely to be co-natal. Nevertheless, these simulations lack direct N-body effects on clusters and individual stars, as well as essential feedback processes.

In this paper, we examine chemical abundance trends in OCs identified in high-resolution cosmological hydrodynamic galaxy simulations generated with the FIRE-2 \citep[Feedback in Realistic Environments,][]{hopkins_2018_fire2} model.
These simulations have an adaptive spatial resolution, which enables modeling a large dynamic range (from Mpc down to pc), and include many relevant star cluster-scale feedback processes.
In this work, we utilize three MW-mass galaxies from the \emph{Latte} suite \citep{wetzel_reconciling_dwarf_galaxies,  wetzel_2023_FIRE2_data_release} of FIRE-2 simulations to assess the viability of strong chemical tagging via inter-cluster and intra-cluster chemical homogeneity analysis.
Importantly, this study is possible now because FIRE-2 implements physically motivated processes (e.g., subgrid turbulent metal diffusion, stellar feedback, and chemical enrichment from core-collapse supernovae, white-dwarf supernovae, and stellar winds) and has high enough resolution in \emph{Latte} (up to $\sim$1 pc) to distinguish individual star-forming regions.

Previously in FIRE-2, 
\citet{bellardini_2021_3d_gas} showed that their MW-mass galaxies have azimuthal scatter in gas at $z = 0$ similar to that observed in nearby galaxies of similar mass.
\citet{matthew_bellardini_2022_3d_abundances_FIRE} investigated abundance trends in all young stars across the \emph{Latte} disks, considering the evolution over time. 
They found that azimuthal variation dominates at $z \gtrsim  0.8$, whereas radial gradients dominate at late times.
Subsequently, \citet{andreia_carrillo_2023_age_metallicity_for_disk_stars_in_MW_simulations} explored distributions in mono-age populations in one \emph{Latte} disk, m12i, as a function of [Fe/H], [X/Fe], birth and present day locations, and found some evidence for inside-out radial growth for stars with ages <7 Gyr. 
\citet{andreia_carrillo_2023_age_metallicity_for_disk_stars_in_MW_simulations} also examined the age--[X/Fe] relation across the disk and found that the qualitative trends agree with observations (apart from C, O, and Ca) with small intrinsic scatter (0.01<$\sigma$<0.04 dex). 
Additionally, \citet{andreia_carrillo_2023_age_metallicity_for_disk_stars_in_MW_simulations} found $\sigma$ to be metallicity dependent, with $\sigma\approx$0.025 dex at [Fe/H]=$-0.25$ dex versus  $\sigma\approx$0.015 dex [Fe/H]=0 dex.
A similar metallicity dependence is seen in the GALAH survey for the elements in common \citep{Sharma2022, andreia_carrillo_2023_age_metallicity_for_disk_stars_in_MW_simulations}. 
Moreover, like \citet{matthew_bellardini_2022_3d_abundances_FIRE}, \citet{andreia_carrillo_2023_age_metallicity_for_disk_stars_in_MW_simulations} found that $\sigma$ is higher in the inner galaxy, where stars are older and formed in less chemically homogeneous environments.

Building on the work of \citet{matthew_bellardini_2022_3d_abundances_FIRE}, \citet{graf_2024_spatial_variation_abundances_current_and_past} recently showed that the radial metallicity gradient as a function of stellar age in the \textit{Latte} disks follows similar trend as in Milky Way, albeit with shallower gradients in FIRE. 
Additionally, \citet{graf_2024_spatial_variation_abundances_current_and_past} showed that at present day the vertical metallicity gradient is steepest for the youngest stellar populations, which is again qualitatively consistent with the Milky Way, despite weaker trends in FIRE.
Finally, both \citet{matthew_bellardini_2022_3d_abundances_FIRE} and \citet{graf_2024_spatial_variation_abundances_current_and_past} considered the azimuthal scatter of [Fe/H] in stars in \textit{Latte}, and found it to be similar to the Milky Way 
\citep[relative to][]{anders_2014_chemodynamics_MW_APOGEE}.

While the FIRE studies discussed above focus on all stars in simulated galactic disks, our focus in this study is on chemical trends in OCs specifically. 
Along this line, in 2023, \citeauthor{grudic_2023_great_balls_FIRE} 
mapped giant molecular clouds formed self-consistently in a \emph{Latte} simulation on to a cluster population according to a giant molecular cloud-scale cluster formation model calibrated to higher resolution simulations.
This approach enabled them to explore the galaxy's star clusters as a function of mass, metallicity, space, and time.
\citet{grudic_2023_great_balls_FIRE} concluded that massive clusters do not form with metallicities differing from other stars forming within the galaxy.

In this work, we take a different approach: we directly identify young massive OCs in the \emph{Latte} simulations at present day to measure their chemical homogeneity in the full galactic context. 
As of yet, a robust comparison between the chemical homogeneity of observed OCs and OCs formed in cosmological Milky Way-mass galaxy simulations does not exist, and we aim to characterize the precision requirements for metallicity measurements to ensure that star cluster reconstruction could be conducted using chemistry alone. 
In this study, we leverage OCs identified within the last $\sim$100 Myr of the simulations, and select OCs younger than 3 Myrs, which allows us to study truly co-forming populations often still embedded in their natal environment. 
We calculate the intra and inter-cluster metallicity dispersion for our OC population and compare it to results from spectroscopic OC observations. 
Finally, we calculate a chemical difference metric between intra-cluster and inter-cluster pairs of stars and discuss the viability of chemical tagging using such chemical differences.

This paper is structured as follows: in \S\ref{sec:data and methodology}, we discuss our data and methodology.  
Specifically, in \S\ref{subsec:sims} we introduce the simulations, in \S\ref{sec 2.2: star_cluster_identification} we discuss how we identify OCs in these  simulations, and in \S\ref{subsec: chemical difference metric} we introduce the chemical difference metric we use to characterize the similarity of elemental abundances. 
In \S\ref{sec:results}, we present our results, and finally, in \S\ref{sec:discussion_conclusions} we present our discussion in context of the observational literature and our conclusions.

\section{Data and Methodology} \label{sec:data and methodology}
\subsection{Simulations}
\label{subsec:sims}

We use the \emph{Latte} suite of FIRE-2 cosmological Milky Way-mass galaxy simulations \citep{wetzel_reconciling_dwarf_galaxies, wetzel_2023_FIRE2_data_release}
to study properties of OCs.
As mentioned earlier, FIRE stands for Feedback in Realistic Environments; the code is able to simulate the interplay between ISM and the stellar feedback processes in a cosmological environment \citep{hopkins_2015_mesh_free_hydrodynamic_simulations, hopkins_2018_fire2}.
As detailed in \citet{hopkins_2015_mesh_free_hydrodynamic_simulations}, these simulations use the Lagrangian Meshless Finite Mass (MFM) hydrodynamics method implemented in Gizmo. 
Gizmo solves the fluid equations using a moving particle distribution that is automatically adaptive to resolution.

FIRE-2 uses a physically motivated metallicity-dependent radiative transfer heating and cooling approach for gas that includes free-free, photoionization and recombination processes, Compton scattering, photoelectric heating and collisional dust effects, as well  as molecular, metal-line, and fine structure processes.
The implementation of FIRE we use in this work additionally includes magnetohydrodynamics, where the equations of ideal magnetohydrodynamics are solved explicitly \citep{Hopkins_2020_MHD}.

We use versions of these simulations that we resimulated over the final 110 Myr specifically to store finer snapshot time spacing (1 Myr) than in the original simulations ($\approx 22$ Myr). This enables us to track OCs in detail in phase space across time. While these resimulations are not publicly available, snapshots of the original simulations are publicly available \citep{wetzel_2023_FIRE2_data_release}.

FIRE-2 tracks 11 elements (H, He, C, N, O, Ne, Mg, Si, S, Ca, Fe) across a temperature range of 10 to $10^{10}$ K. 
These simulations incorporate a spatially uniform, redshift-dependent UV background as described by \citet{faucher_2009_ionizing_background_spectrum}. 
As typically done by observers, metallicity and elemental abundances are scaled to solar values; in these simulations, we use values for the Sun from \citet{asplund_2009_chemical_composition_of_sun}.

Star formation proceeds according to the following recipe in these simulations: each star particle is generated from a self-gravitating gas cell that is Jeans-unstable, cold ($T < 10^{4}$ K), dense (n $>$ 1000 $cm^{-3}$), and molecular \citep[following][]{krumholz_2011_molecular_content_of_model_galaxies}; when star formation criteria are met, there is a one-to-one conversion of a gas cell into a star particle. 
That is, a star particle acquires its mass and elemental makeup from the gas cell it originates from and depicts a simple stellar population characterized by a Kroupa initial mass function \citep{kroupa_2001_variation_of_the_initial_mass_function}.
Each star particle is individually evolved assuming a standard stellar population model \citep[STARBURST99 v7.0,][]{leitherer_1999_starburst99} and produces time-resolved stellar feedback from both core-collapse and white-dwarf supernovae. The rates of core-collapse supernovae are based on STARBURST99 \citep[][]{leitherer_1999_starburst99} and the rates of white-dwarf supernovae are from \citep{mannucci_2006_progenitors_for_type_1a_supernovae}.
FIRE-2 also incorporates stellar feedback due to continuous mass loss, radiation pressure, photoionization, and photoelectric heating.
The nucleosynthetic yields we use are based on \citet{iwamoto_1999_nucleosynthesis} for white-dwarf supernovae and \citet{nomoto_2006_nucleosynthesis_yeilds_of_core_collapse_supernovae} for core-collapse supernovae.
Stellar wind yields are sourced primarily from O, B, and AGB stars; we use a model for this assembled by \citet{wiersma_2009_chemical_enrinchment_in_simulations} which incorporates expectations from \citet{van_den_hoek_1997_theoretical_yields_of_intermediate_mass_stars}; \citet{marigo_2001_chemical_yields}; \citet{izzard_2004_synthetic_model_giant_branch_stars}.
Further details on this can be found in Appendix A from \citet{andreia_carrillo_2023_age_metallicity_for_disk_stars_in_MW_simulations} and in \citet{hopkins_2018_fire2}.

FIRE-2 incorporates sub-grid diffusion and mixing of elemental abundances in gas, which occurs through unresolved turbulent eddies \citep{su_2017_feedback_first,escala_2018_modelling_chemical_abundance,hopkins_2018_fire2}.
Our model assumes sub-grid mixing is primarily influenced by the largest unresolved eddies, which effectively smooths out abundance variations among gas elements.
\citet{escala_2018_modelling_chemical_abundance} demonstrated that incorporating this sub-grid model is essential for accurately replicating observed distributions of stellar metallicities in galaxies.
\citet{bellardini_2021_3d_gas} showed that while the details of this diffusion model do not significantly affect radial or vertical gradients in FIRE-2 MW-mass galaxies, they do affect the azimuthal scatter.
\citet{bellardini_2021_3d_gas} also showed that these galaxies have azimuthal scatter in gas at $z = 0$ similar to that observed in nearby galaxies of similar mass.

The \emph{Latte} suite contains several Milky Way-like galaxies;  
in this work, we analyze three  \emph{Latte} simulations (m12f, m12i, and m12m).
The galaxies we analyze represent some of the most massive Milky Way-like isolate disks in the \emph{Latte} suite and span a range of assembly histories. 
Key differences among these simulations are that m12i has a massive disk with a late-forming Milky Way mass halo, m12f has an LMC-like satellite merger with the Milky Way-like disk 3 Gyr prior to present day, and m12m has a strong bar at high redshift and an earlier forming halo \citep[see][]{sanderson_2020_ananke}. 
A full description of the criteria used to generate these galaxies and their properties at the present day can be found at \citet{wetzel_2023_FIRE2_data_release} and references therein. 
A brief description of relevant details to this work is as follows.
Each galaxy simulation assumes a flat $\Lambda$CDM cosmology with parameters consistent with the \citet{plank_collaboration_2020}: $h = 0.68 - 0.71$, $\Omega_{\Lambda} = 0.69 - 0.734$,
$\Omega_{m} = 0.266 - 0.31$, $\Omega_{b} = 0.0455 - 0.048$, $\sigma_{8}= 0.801 - 0.82$ and $n_{s} = 0.961 - 0.97$. 
\footnote{Where $h$ is the dimensionless equivalent of the Hubble Parameter $H_{0}$, $\Omega_{\Lambda}$ is the cosmological constant or vacuum density at present time, $\Omega_{m}$ is the total matter (dark plus baryonic) density today, $\Omega_{b}$ is baryonic matter density, $\sigma_{8}$ is the root mean squared fluctuation of density perturbations at 8 $h^{-1}Mpc$ scale and $n_{s}$ is the scalar spectral index.}

In this work, Milky Way-like refers to the total mass of each galaxy at present day. 
The \emph{Latte} galaxies were selected to have halo masses roughly the same as the Milky Way's: $M_{200} = 1 $--$ 2 \times 10^{12} M_{\odot}$ where $M_{200}$ refers to the total mass within the radius containing 200 times the mean matter density of the Universe.
These simulations contain three types of material: dark matter, stars, and gas. The resolution of each dark matter particle is $3.5 \times 10^{4} M_{\odot}$ and the initial mass resolution of baryonic material  is $7070\ M_{\odot}$. 
As noted earlier, each star particle should be thought of as a single stellar population. Because of this, as a star particle ages, it loses a portion of its stellar mass. 
Thus, at $z = 0$ a typical star particle has a mass of $\sim 5000\ M_{\odot}$.
In the \emph{Latte} galaxies, star and dark matter particles have a fixed gravitational force softening length of 4 and 40 pc (Plummer equivalent) respectively.
Gas cells have adaptive gas smoothing: in regions where there is a high density of gas cells, the gravitational force softening length can resolve down to a minimum of 1 pc.

Given its spatial and mass resolution, the \emph{Latte} simulations can resolve massive star clusters with multiple star particles that represent portions of each OC. 
And, as mentioned earlier, with the temporal resolution for snapshots of 1 Myr for these re-simulations, they can track in detail how star particles in OCs evolve in phase space over time.
See \S\ref{sec 2.2: star_cluster_identification} for more details on OC selection and characteristics.

\begin{figure}
    \includegraphics[width=\columnwidth]{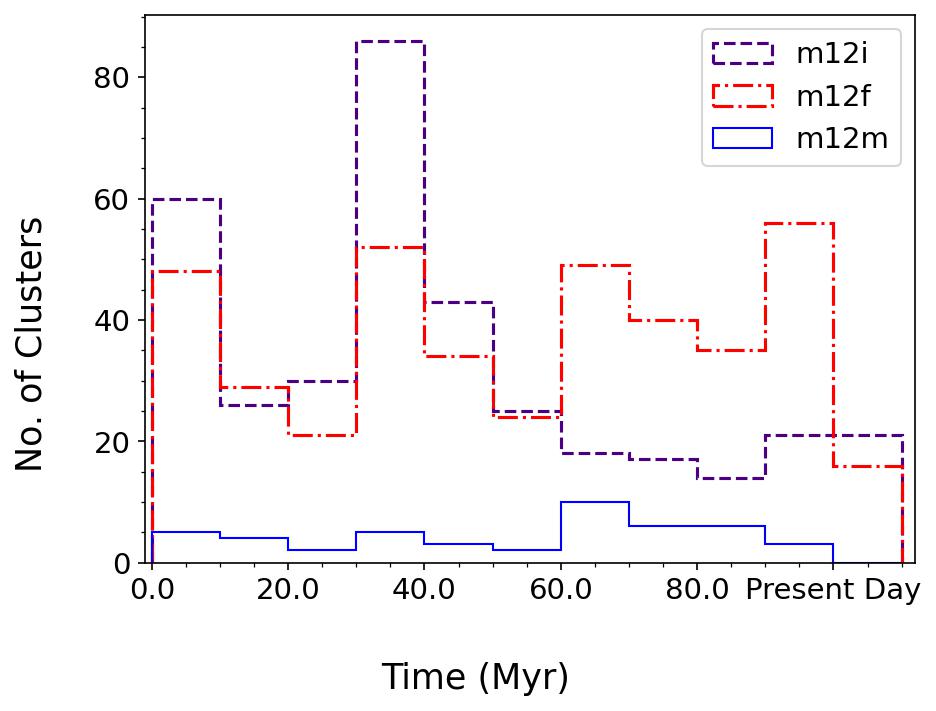}
    \caption{Number of young (<3 Myr) massive OCs identified in the last $\sim$100 Myr of the simulations. 
    OC selection has been run at a 3 Myr interval to avoid double-counting. 
    m12i indicated by dashed lines colored in indigo, m12f indicated by dashed-dotted lines colored in red, and m12m indicated by solid blue lines.
    A varying rate of star cluster formation is observed in all three simulations, with m12m having the lowest number of clusters produced throughout the 100 Myr period. This reflects a globally suppressed star formation rate for m12m at present day. See \S~\ref{sec 2.2: star_cluster_identification} for further discussion.}
    \label{fig: No. clusters vs time}
\end{figure}

\begin{figure}
    \includegraphics[width=\columnwidth]{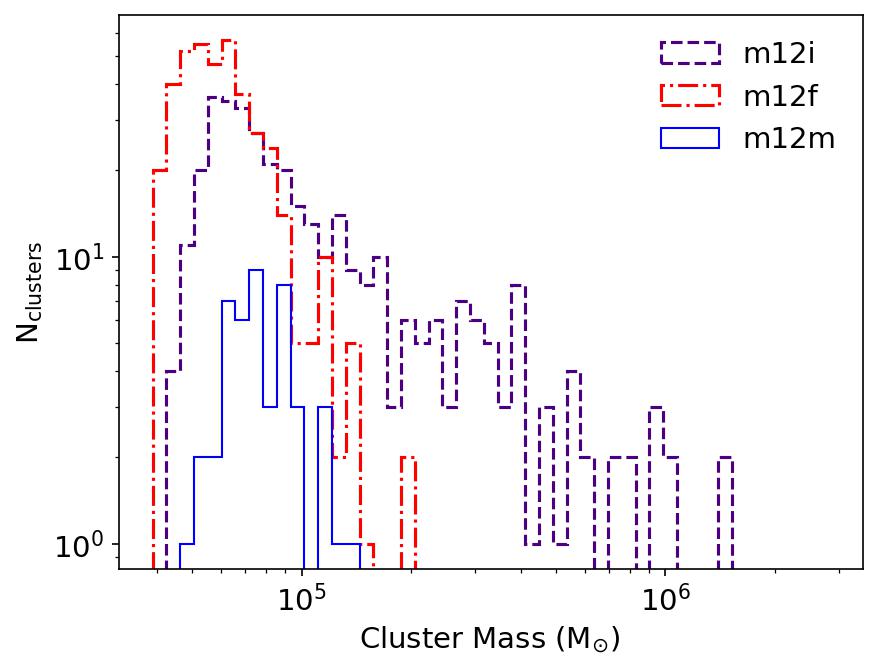}
    \caption{Mass distribution of the OCs from all three simulations. m12i indicated by dashed lines colored in indigo, m12f indicated by dashed-dotted lines colored in red, and m12m indicated by solid blue lines. The slope of the cluster mass function is similar for all three simulations with m12i having a shallower slope with a few high-mass outliers.}   
    \label{fig: No. of clusters with a given mass histogram}
\end{figure}

\begin{figure*}
    \includegraphics[width=\textwidth]{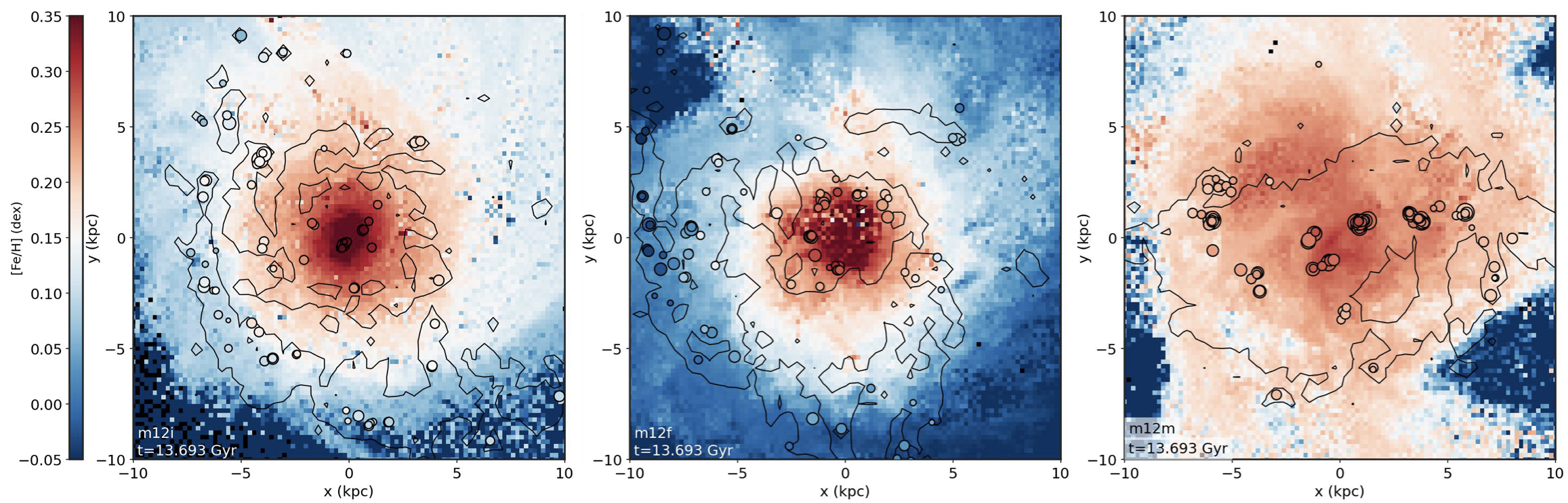}
    \caption[Snapshots from three \emph{Latte} simulation]{A snapshot from three \emph{Latte} simulations m12i, m12f and m12m. Background: mean [Fe/H] for gas in the disk of each simulation. The corresponding values are shown by the color bar on the left. Contours: density of stars $<$ 50 Myr old which highlights the spiral arms. Circles: mean [Fe/H] of OCs and associations identified at 13.693 Gyr.
    Note, the mean [Fe/H] of the clustered star formation reflects the global metallicity distribution of the gas in the disk.}
    \label{fig:Background [Fe/H] distribution m12i}
\end{figure*}

\begin{figure}
    \includegraphics[width=\columnwidth]{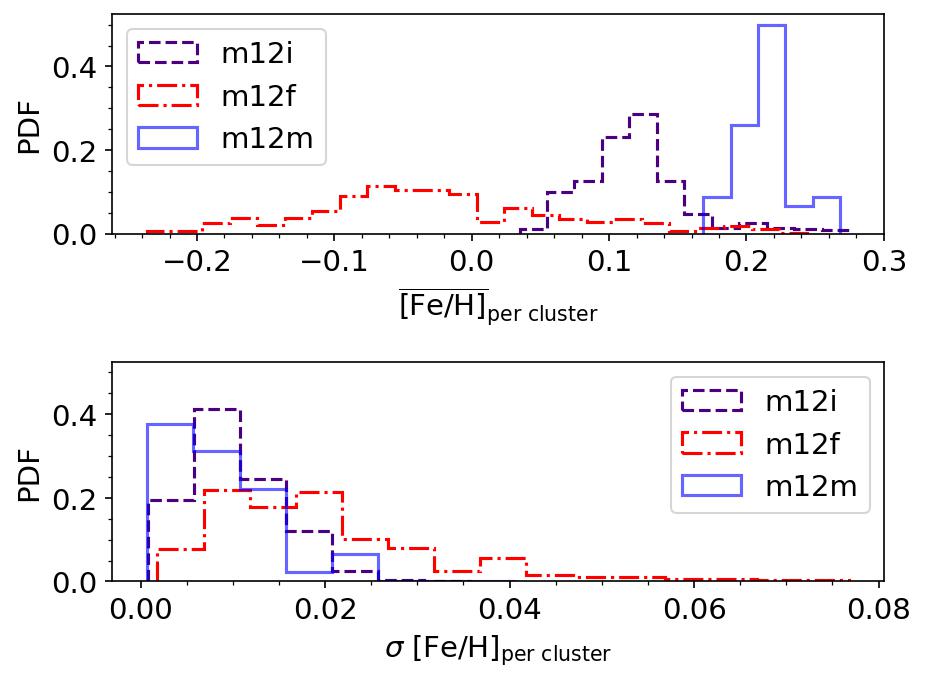}  
    \caption{Distribution of mean [Fe/H] (upper panel) and dispersion of [Fe/H] in the cluster sample of all three simulations. m12i indicated by dashed lines colored in indigo, m12f indicated by dashed-dotted lines colored in red, and m12m indicated by solid blue lines. m12f has a larger range of mean and scatter in [Fe/H].}     
    \label{fig: mean feh and dispersion all sims}
\end{figure}

\subsection{Star Cluster Identification}
\label{sec 2.2: star_cluster_identification}
 We identify OCs in our simulations using the friends-of-friends (FoF) algorithm \citep{geller_1983_fof_citation}. 
It works by identifying members within a fixed distance known as the linking length; in this work we adopt a linking length of 4 pc, which corresponds to our softening length.
To ensure that the OCs are identified before the onset of core-collapse supernovae, we select OCs from star particles that are less than 3 Myr old over the last 110 Myr of these simulations. 
We require a minimum of 5 star particles for a cluster to be identified, and thus with our fiducial mass resolution (7070 $\mathrm{M}_{\odot}$) we resolve cluster down to a cluster mass of $10^{4.6}$ $\mathrm{M}_{\odot}$.

With these selection criteria, the OCs we identify have a characteristically small velocity dispersion ($\sigma_{1D}$$\sim$$4$ km/s), small size ($R_{\mathrm{half}}$$\sim$$3$ pc), and small spread in age ($\sigma_{\mathrm{age}}$$\sim$$.25$ Myr). 
We do not require that the natal giant molecular cloud be bound for our OCs to  form inside them; despite this, 
roughly half of the clusters that we identify are born bound ($\log_{10}(2KE/PE) \ge 0$) and the global population follows a Gaussian distribution of boundedness peaked at 0.
While the size and boundedness of the population that we identify may be better described as OCs and associations, for the purposes of this study, which aims to characterize the spread in metallicity inside and between small clustered star forming environments across a full galactic disk, we find it sufficient not to distinguish between OCs and associations.
We also note that the age selection criteria that we adopted does not significantly impact any of these properties \citep{Masoumi2024}.
Because of this, we feel that the OC populations that we identify sufficiently resemble OCs within a full simulated cosmological Milky Way-like galaxy environment to proceed with the elemental abundance study we present here; a more detailed study that focuses on the impact of the selection crieria we use on the resultant properties of the OCs we identify can be found in \citet{Masoumi2024}.
Note we identify star clusters similarly in all three Milky Way-like analog simulations mentioned in \ref{subsec:sims}.

In this work, we identify a total of 361, 404, and 46 unique OCs in m12i, m12f, and m12m respectively using the criteria discussed above.
We analyze the OCs from each galaxy simulation as a single dataset.
In Fig. \ref{fig: No. clusters vs time} we show a histogram of the number of clusters identified throughout the final 110 Myr of the simulations. 
These 3 simulations (m12i, m12f, and m12m) represent a varying rate of star cluster formation over the 110 Myr epoch; m12m has a relatively consistent albeit low number of clusters formed over this period of time, while m12i forms roughly a quarter of all the clusters over just a 3 Myr span, and finally m12f has a fluctuating rate of cluster formation that oscillates on a $\sim$30 Myr timescale.  
We note that the low rate of m12m's cluster formation reflects the overall depressed star formation rate of the entire galaxy at the present day.

The cluster mass distribution of these simulations is shown in Fig.~\ref{fig: No. of clusters with a given mass histogram}. 
We note that m12f and m12m have similar slopes for the cluster mass function. 
However, m12i has a shallower slope and contains a few high-mass outliers that form during a period of vigorous star formation in the plane of the disk. 
These OCs are best thought of as Young Massive Clusters (YMCs) and will be discussed further in Masoumi et al. (in prep).

In Fig.~\ref{fig:Background [Fe/H] distribution m12i}, we show a top-down view of m12i, m12f, and m12m near the present-day (13.693 Gyr). 
Here we show gas in the background which has been binned in 200 pc x 200 pc squares and colored by the mean value of [Fe/H] in each volume.
Plotted on top of the average [Fe/H] distribution for gas are contours of the density distribution of stars that are less than 50 Myr old. 
These contours highlight where there is an active star formation corresponding to the spiral structure. 
The circles overlayed on top indicate the mean [Fe/H] of OCs and associations (linking length $\leq$18 pc) that were identified at this moment in time; while we have included larger associations here for visualization purposes, in this paper, we only include OCs found using a linking length of 4 pc for analysis.
We note that Fig.~\ref{fig:Background [Fe/H] distribution m12i} shows that the average [Fe/H] of OCs and associations generally reflects the gas metallicity distribution across the disk.

We note that the top-down perspective on the galactic disks shown in Fig.~\ref{fig:Background [Fe/H] distribution m12i} highlights the overall radial metallicity gradient in all three galaxies; while each galaxy has a different normalization and slope of the radial metallicity gradient, overall it is clear that a negative radial gradient is present in the global disk for gas at present-day \citep[see][for further discussion]{bellardini_2021_3d_gas}.
While here we have shown the distribution of [Fe/H], we will be using a total of 9 elemental abundances (Fe, C, N, O, Mg, Si, S, Ca, Ne) throughout the remainder of this work. 
We note that these elements are highly correlated with each other in our simulations. 
We show this correlation in Appendix~\ref{sec:Appendix_correlation_between_elements} in Fig.~\ref{fig: correlation between element over feh} and Fig.~\ref{fig: correlation between element over another element}. 
Such a strong correlation is also seen in the Milky Way \citep{griffith_2024_KPM};  in fact, \citet{ness_2022_homogeneity_of_star_forming_environment_in_the_milky_way} show that, using a linear regression fit to Fe and Mg alone, one can predict eight supernova elements within 5\% accuracy.
Despite such strong correlations, we elect to use all elements at our disposal in this work. 

In Fig.~\ref{fig: mean feh and dispersion all sims}, we show
normalized histograms of the mean [Fe/H] (upper panel) and standard deviation of [Fe/H] (lower panel) for OCs in all three simulations.
The upper panel shows the presence of scatter in mean [Fe/H]. 
While m12f has a large range of OC mean [Fe/H], m12i and m12m have a narrower range of values.
Each of these populations peaks at a different value of [Fe/H]; however, there is some overlap between these samples.
The standard deviation histogram for each galaxy shown in the bottom panel of  Fig.~\ref{fig: mean feh and dispersion all sims} illustrates that there are very few OCs that have a large standard deviation of [Fe/H]; moreover, the peak of each distribution is at a small value. 
Given these essential properties -- there is scatter in average [Fe/H] in OCs in the simulations and there is small spread in [Fe/H] in each OC -- in this work, we are able to test the feasibility of uniquely identifying stars from the same OC using their elemental abundances alone.

\subsection{Chemical Difference Metric} \label{subsec: chemical difference metric}

The goal of this investigation is to understand the similarity of stellar abundances for stars within a given OC; to simplify this, we define a simple chemical difference metric using a pseudo-Cartesian measurement of distance in abundance space.
A similar approach is used in \citet{ness_2018_doppelgangers},
which was adapted from \citet{mitschang_2013_distance_metric}. 
The metric is defined as:

\begin{center}
$d_{n n^{\prime}}^{2}=\sum_{i=1}^{I} {\left[x_{n i}-x_{n^{\prime} i}\right]^{2}}$
\end{center}

 Where $n$ and $n^{\prime}$ are two stars being compared, with star $n$ containing the abundance $x_{n i}$ and star $n^{\prime}$ containing the abundance $x_{n^{\prime} i}$ for all elements 1 to $I$. 
Simply put, this metric evaluates the squared chemical difference between any two stars, summed over all available elements, and provides a single number that accesses the chemical similarity for each pair of stars. 
In this work we initially evaluate the chemical difference metric summed over 8 element abundances ([Mg/Fe], [Ca/Fe], [S/Fe], [Si/Fe], [Ne/Fe], [O/Fe], [N/Fe], [C/Fe]) without any errors being considered.

 After this initial assessment, we further consider the chemical difference metric with APOGEE errors added to our simulated OCs \citep[drawn from  Chemical Abundances and Mapping Survey,][]{natalie_myers_2022_OCCAM}. 
 The difference metric with added errors is defined as:

\begin{center}
$d_{n n^{\prime}}^{2}=\sum_{i=1}^{I} \frac{\left[x_{n i}-x_{n^{\prime} i}\right]^{2}}{\sigma_{n i}^2+\sigma_{n^{\prime} i}^2}$
\end{center}

Where $\sigma_{n i}$ refers to the error measurement for $x_{n i}$ and $\sigma_{n^ {\prime} i}$ represents the error measurement for $x_{n^{\prime} i}$. 
Note we omit Ne (neon) from the above list of elements in this analysis as it is a relatively rare element that isn't measured in APOGEE OCs.

\section{Results} \label{sec:results}

\subsection{The relationship between [Fe/H] $\&$ birth radius for OCs}
\label{subsec: feh vs birth radius}

In this section, we consider the radial metallicity gradient: the average [Fe/H] as a function of galactocentric radius, which is often thought to be a unique indicator of where stars are born \citep[e.g.,][]{Minchev_2018}. Here we measure the radial metallicity gradient using OCs, as has often been done in the literature \citep{janes1979evidence,sestito_2008_OC_key_tracers,spina_2022_challenges_OCs,magrini_2023_abundance_gradient_OCs}.
Our goal is to assess how truly unique mean [Fe/H] is for OCs in our simulations.
We note that the dispersion in each cluster does not change substantially across the disk.

\begin{figure}
    \includegraphics[width=\columnwidth]{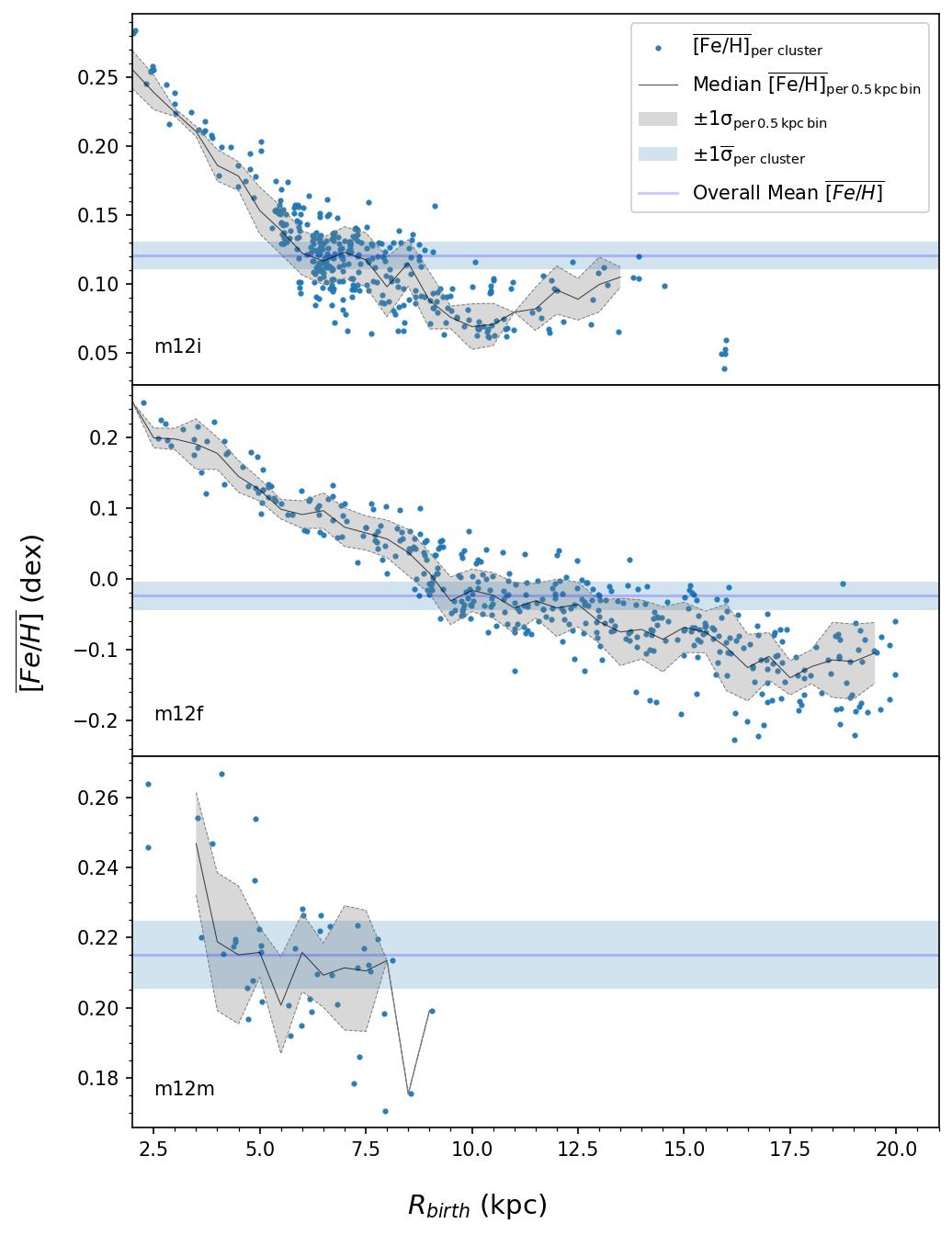}
    \caption{Scatter plot of mean [Fe/H] vs birth radius of OCs from 3 simulations: m12i (top panel), m12f (middle panel), m12m (bottom panel). The blue dots indicate the mean [Fe/H] of each cluster and the solid black curve represents the median trend for data in 0.5 kpc radial bins. The grey shaded area represents the $\pm{1 \sigma}$ range within 0.5 kpc bins. The blue horizontal line represents the average of all OCs in the sample, and the blue horizontal shaded region represents the typical $\pm{1\sigma}$ scatter of [Fe/H] within any given cluster. It can be seen that the metallicity distribution varies differently in the inner disk than the outer, and many clusters have similar mean [Fe/H] despite being found at varying locations across the disk.}
    \label{fig:feh vs R_CM}
\end{figure}

In Fig. \ref{fig:feh vs R_CM}, we show the mean [Fe/H] vs the birth radius ($\mathrm{R_{birth}}$) for all OCs identified over the final 110 Myr in m12i (top panel), m12f (middle panel), and m12m (bottom panel). 
Note that the range of values indicated in y-axes are slightly different for each galaxy, which is very important while comparing slopes between each system.
Each blue dot indicates the mean [Fe/H] value of an individual OC. 
We calculate a running median using a bin size of 0.5 kpc (shown in a black solid line) and indicate the $\pm 1 \sigma$ confidence region around the median values, shaded in grey. 
In addition to this, we indicate on each figure the global average [Fe/H] (solid blue horizontal line) of all our OCs and also the average dispersion of [Fe/H] (shaded blue region) inside a typical OC. 
We later subsample OCs drawn from this blue-shaded region to perform a chemical difference metric test (see section \ref{3.3 chem diff metric analysis}).

We find a very different trend in the metallicity gradient in the three galaxies we assess. For m12i (top panel), there is a steep gradient within $\mathrm{R_{birth}}$ $<$ 6 kpc (with a slope of $\mathrm{\sim -0.02\ dex\ kpc^{-1}}$) and a small scatter about the median trend line (with scatter $\mathrm{< 0.01 \ dex}$). 
This combination of steep slope and small scatter means that, within this region, [Fe/H] strongly constrains birth location. 
However, at $\mathrm{R_{birth}}$ $>$ 6 kpc, the slope of the radial metallicity gradient substantially flattens out and the scatter significantly increases ($\mathrm{> 0.02 \ dex}$). Moreover, outside of 10 kpc, there is an inversion in the slope of the metallicity gradient. Taken together, there is no longer a one-to-one mapping of [Fe/H] and birth radius outside 6 kpc and [Fe/H] should not be taken as a reliable indicator of the birth location.

However, when we consider m12f (middle panel), the predictive power of [Fe/H] as a unique indicator of birth location extends to a significantly larger region (out to 10 kpc). 
Within this radial range, there is a small scatter around the median trend line (< 0.02 dex).  
Moreover, like m12i, within the inner region of m12f, the slope has a constant negative value of $\mathrm{-0.04\ dex\ kpc^{-1}}$.
It is noteworthy that, for m12f, outside of 10 kpc the scatter increases significantly, doubling to $\sim$0.04 dex in the the outer regions of the disk. 
And, like m12i, at the largest Galactocentric birth radii, there is a subtle upturn in the median trend line.
Thus, again, at large Galactocentric radii, [Fe/H] no longer has strong predictive power as an indicator of birth location.

Finally, m12m (lower panel), has a significantly smaller number of clusters than the other two systems, and these clusters are all found at Galactocentric radii < 9 kpc.  
While the slope of the radial metallicity gradient for m12m is negative, it is substantially shallower ($\mathrm{< -0.008\ dex\ kpc^{-1}}$) than m12i and m12f.
Moreover, while the typical scatter about the median trend line is similar to the inner region of m12i ($\sim$0.01 dex), in combination with the relatively shallow gradient, there is limited discerning power for [Fe/H] to indicate birth location in this system.  

Overall all three systems appear to have a negative gradient at small radii that generally flattens at larger radii consistent with the analysis of these galaxies in \citet{matthew_bellardini_2022_3d_abundances_FIRE}. 
Such a gradient and flattening in the average metallicity of OCs as a function of radius has often been modeled in the Milky Way using a two-component fit \citep[e.g.,][]{donor_2020_OC_chemical_abundances,spina_2022_challenges_OCs,natalie_myers_2022_OCCAM}. 
The presence of such a transition in the gradient is an indicator that variation in abundances is different for different regions of our Galactic disk. 

Note, in section \ref{3.3 chem diff metric analysis} we will discuss the shaded blue region in Fig. \ref{fig:feh vs R_CM} and further consider the additional power of individual abundances to draw connections to birth location. 
This can be done by selecting OCs that fall within $\pm{1\sigma}$ of a fixed [Fe/H] value. 
As we mentioned earlier, $\sigma$ is set by the typical dispersion within an OC, while the fixed [Fe/H] value is set by the average [Fe/H] of all the OCs discussed here.

\begin{figure}
    \includegraphics[width=\columnwidth]{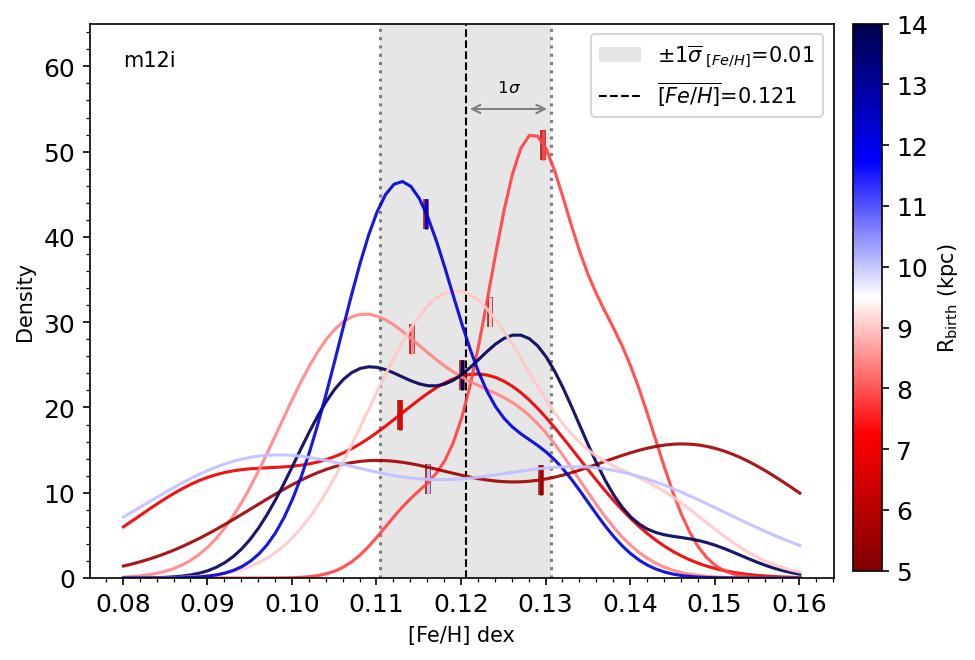}
    \includegraphics[width=\columnwidth]{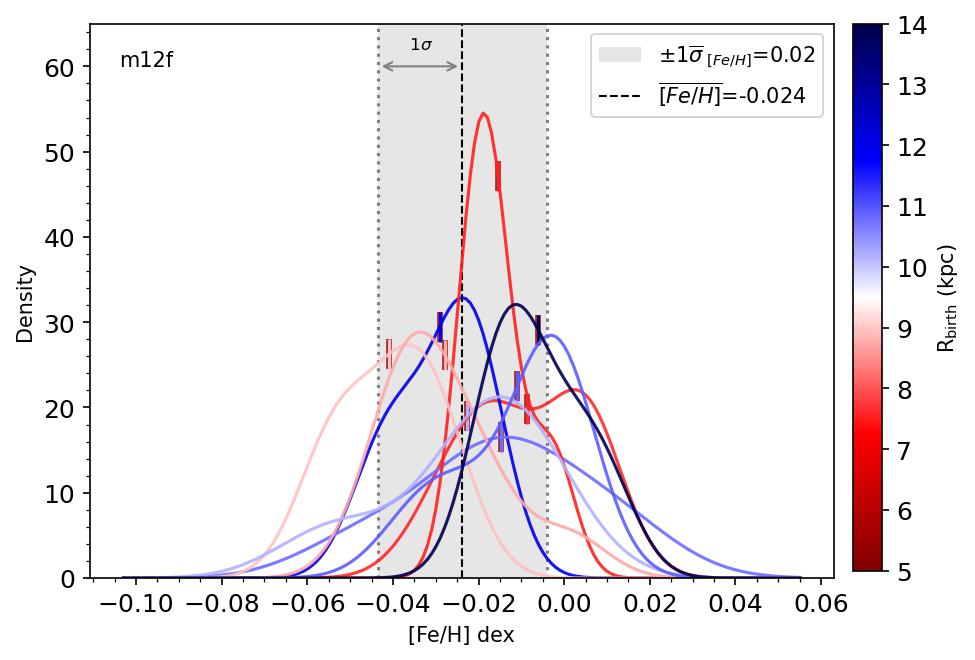}
    \includegraphics[width=\columnwidth]{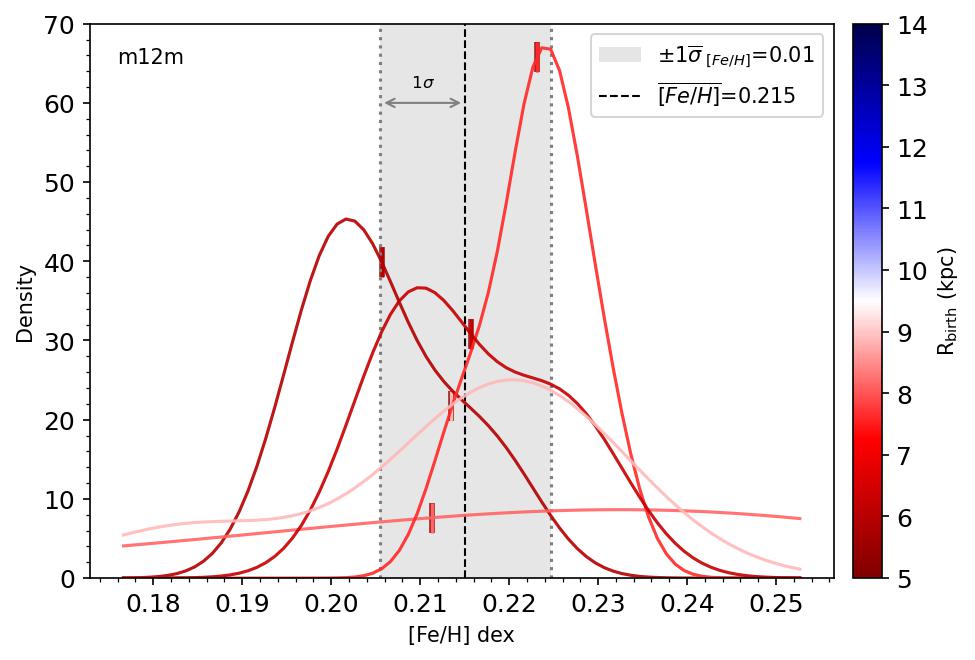}
    \caption{The [Fe/H] distribution for a subset of OCs from the subsample highlighted in blue from Fig.~\ref{fig:feh vs R_CM} for each simulation. Each smoothed histogram represents the metallicity distribution for a single OC and was generated using Gaussian kernel smoothing. The shaded region indicates the $\pm{1\sigma}$ from the overall mean of the distribution of all clusters from that simulation. Despite their locations in the disk, it is possible to find OCs that have similar abundance distributions at different R.}
    \label{fig:6 Fe/H dist for a subset of OCs}
\end{figure}

In Fig.~\ref{fig:6 Fe/H dist for a subset of OCs}, we show the metallicity distribution function for a subset of OCs found within the fixed [Fe/H] range discussed above. These histograms were generated using a Gaussian kernel smoothing technique. Each curve is color-coded by where the OCs formed in the disk. The density shown in the y-axis is calculated such that the area under the curve is 1 using a bin width of 0.001. The black dashed line along the center replicates the horizontal line in Fig.~\ref{fig:feh vs R_CM} which represents the global mean [Fe/H] for all clusters in a given simulation, and the grey-shaded region inside the black dotted lines indicate the 1$\sigma$ dispersion from the typical standard deviation inside each OC. 

The clusters shown in Fig.~\ref{fig:6 Fe/H dist for a subset of OCs} represent a subset of all clusters shown in the blue shaded region of Fig.~\ref{fig:feh vs R_CM} and probe a large range of $\mathrm{R_{birth}}$ across the disk. 
The short vertical line on each curve indicates the mean value of [Fe/H] for that OC. 
Here we can see that while these clusters have similar mean [Fe/H] (all fall within the grey-shaded region), their individual metallicity distribution function can vary in shape.
While these clusters form at a very large range in radii, their metallicity distribution function does not map uniquely onto their formation location. 
Notably, the mean [Fe/H] of a young OC does not always map strictly onto the formation location nor does the dispersion of [Fe/H] inside a given OC. 
This reinforces the idea that the mean metallicity may not be enough to uniquely determine the formation locations of OCs.

\subsection{Inter and Intra Cluster Metallicity Dispersion}

The intra-cluster dispersion provides a measure of the chemical homogeneity of member stars within a given cluster for a given element. 
If such a measurement is small, it indicates a similarity in the elemental abundance of member stars inside a cluster. 
Here we define the intra-cluster dispersion as the weighted mean of the dispersion inside each cluster across all OCs for each simulation. 
Here the weight is proportional to the number of star particles in each cluster.

Similarly, inter-cluster dispersion can provide a measure of the scatter in the mean elemental abundance between different clusters. 
A large value of inter-cluster dispersion indicates a significant difference in the mean elemental abundance (e.g., $\mathrm{\overline{[Fe/H]}}$) between OCs. 
One might expect the inter-cluster dispersion to be higher than the intra-cluster dispersion especially when comparing clusters across a large range of radii. 
In this limit, one might expect strong chemical tagging to be feasible. With a larger dispersion between clusters and a smaller dispersion inside a cluster, it would be easier to identify stars that belong to the same .
However, as we saw in \S \ref{subsec: feh vs birth radius}, regions of the disk can have a weak radial metallicity gradient. Because of this, it is valuable to quantify the comparative significance of the intra versus inter-cluster dispersion to test the feasibility of strong chemical tagging.
We will further discuss the comparative significance of these dispersions utilizing a chemical difference metric in the next section.

\begin{figure}
    \includegraphics[width=\columnwidth]{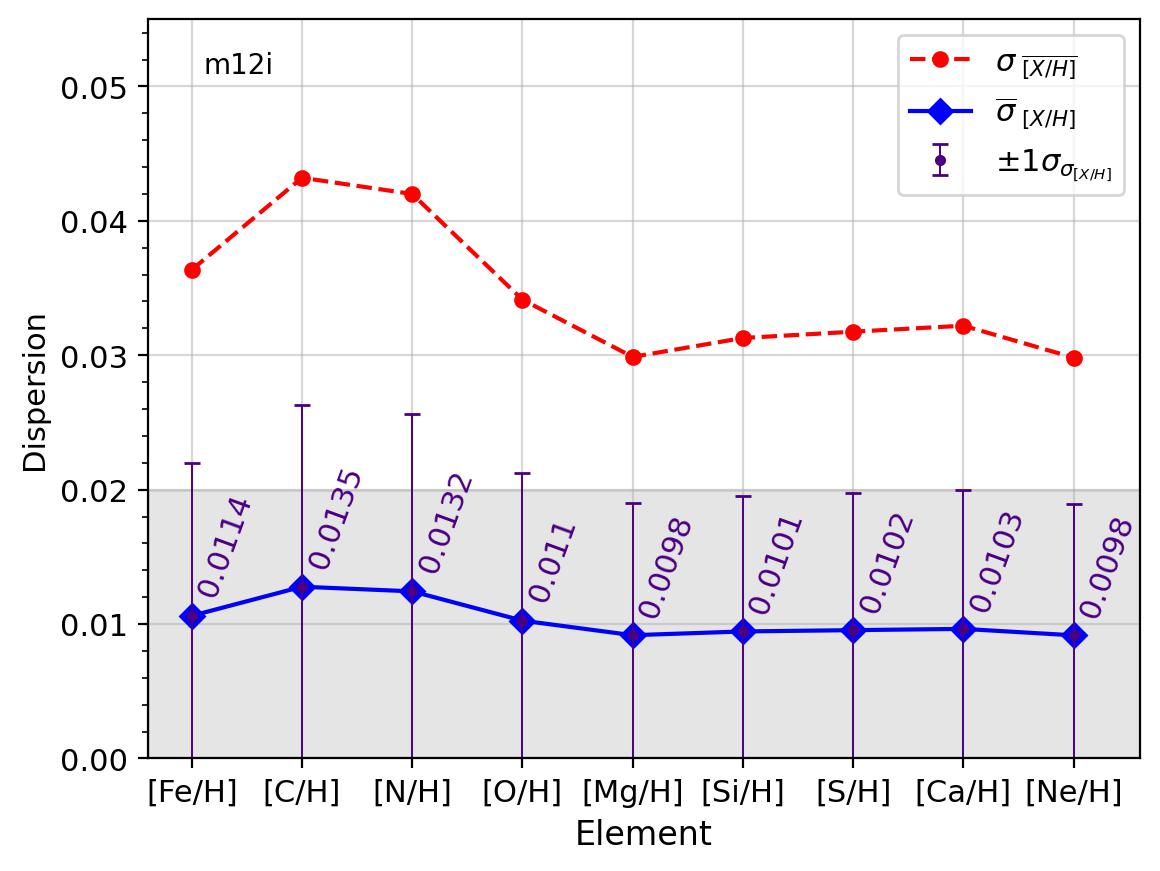}
    \includegraphics[width=\columnwidth]{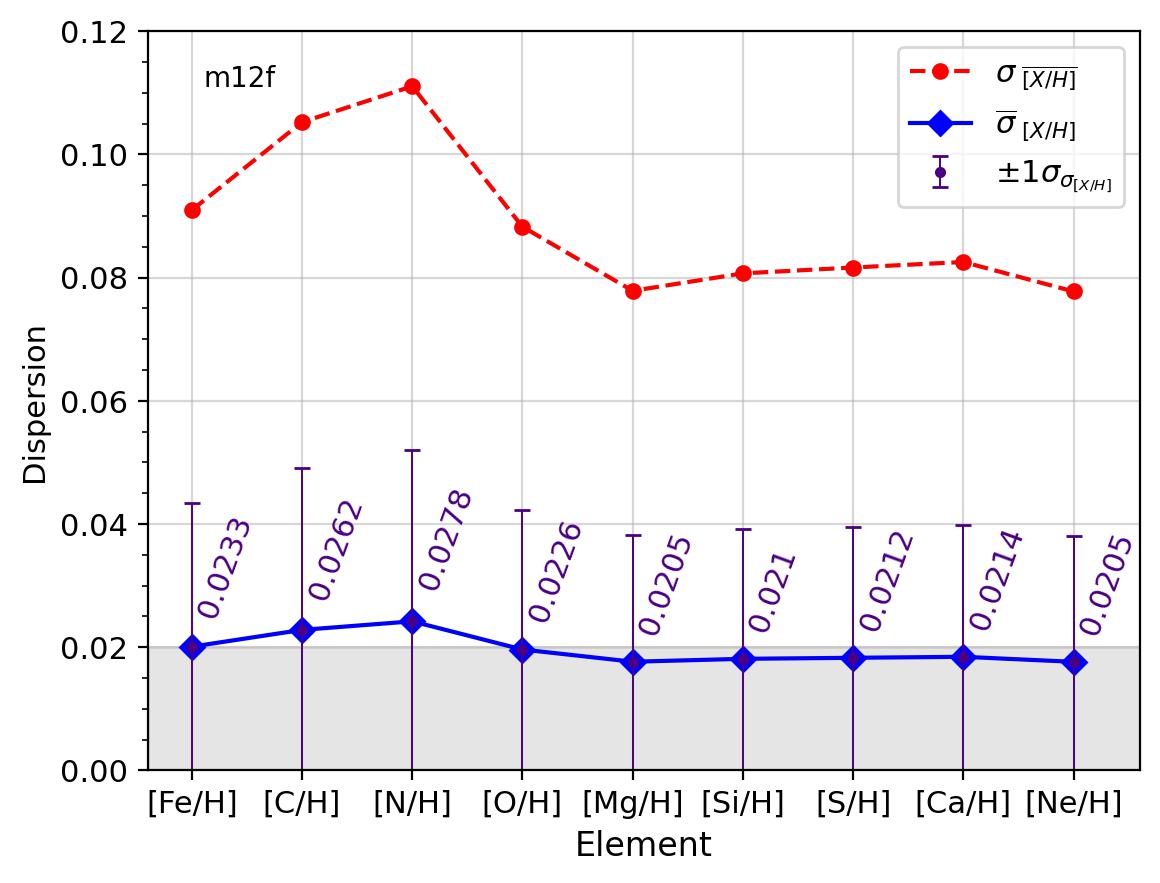}
    \includegraphics[width=\columnwidth]{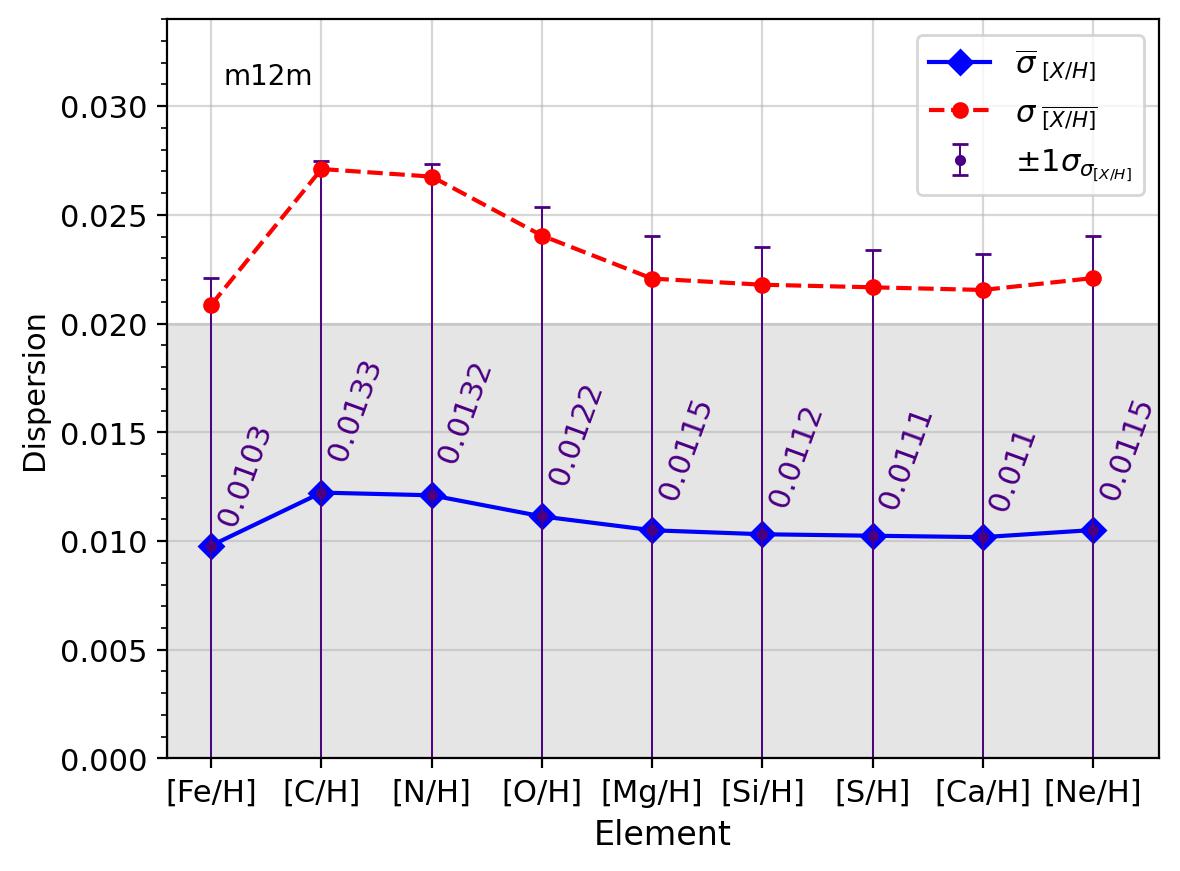}   \caption{For each of the simulations in each panel, Blue: Intra cluster dispersion indicated as the mean dispersion per element. Indigo: the dispersion around average dispersion. The indigo splines indicate the $\pm{1\sigma}$ around it. The grey-shaded region highlights the typical measurements of such dispersion from observations with a grey horizontal line indicating the upper limit. Dashed Red: Inter cluster dispersion. m12f has larger dispersion in both inter and intra-cluster abundances most probably due to an LMC-like merger resulting in a more mixed disk compared to other simulations that don't incorporate such a merger.}
    \label{fig:5 intra and inter cluster dispersion}
\end{figure}

In Fig.~\ref{fig:5 intra and inter cluster dispersion}, we show the intra and inter-cluster dispersion for all OCs in each simulation for each of the 9 elements that we trace (Fe, C, N, O, Mg, Si, S, Ca, and Ne). M12i, m12f, and m12m are shown in the top, middle, and bottom panels respectively; within each panel, the lower curve in blue indicates the weighted average intra-cluster dispersion per element, hereafter represented by the symbol $\overline{{\sigma}}_{~[X/H]}$. 
The vertical indigo lines corresponding to each element indicate the $\pm 1 \sigma$ scatter of the intra-cluster dispersion in all clusters. Since this is a dispersion of the dispersion, it was calculated by first converting these standard deviations into variances. 
For \( n \) standard deviations: \( \sigma_1, \sigma_2, \ldots, \sigma_n \), the variance of the sample standard deviations can be calculated as:
\begin{center}
$\mathrm{{Var}(\sigma_i) = \sigma_i^2}$
\end{center}
We then calculate the average of these variances as:
\begin{center}
$\mathrm{{Mean}(\text{Var}) = \frac{1}{n} \sum_{i=1}^{n} \text{Var}(\sigma_i)}$
\end{center}
\noindent{Finally, the standard deviation of the standard deviation is:}
\begin{center}
    $\mathrm{{Standard\:Deviation}(\sigma_{\sigma}) = \sqrt{\text{Mean}(\text{Var})}}$
\end{center}
The typical measurements of the intrinsic dispersion from observations for these elements fall under the shaded grey region \citep[drawn from ][]{ness_2018_doppelgangers}. The upper red curve indicates the measurement of inter-cluster dispersion (hereafter represented by the symbol ${\sigma}_{~\overline{[X/H]}}$). This has been calculated as the weighted standard deviation of the mean elemental abundance, $\mathrm{\overline{[X/H]}}$, in each cluster.

As indicated by these measurements, the average intra-cluster dispersion inside each FIRE OC is quite small and generally less than 0.020 dex; these measurements are comparable to the intrinsic dispersion calculated by several observational studies, which find OCs have intrinsic dispersion $<$0.03 dex \citep[e.g.,][]{jo_bovy_2016_chemical_homogeneity_of_open_clusters,ness_2018_doppelgangers,bertran_de_lis_2016_variance_ofe,poovelil_2020_oc_chemical_homogeneity_MW}.
This is also consistent with previous studies of m12i, which has considered the global dispersion for stars in the disk at present day.
As mentioned in \S~\ref{sec:introduction}, 
\citet{andreia_carrillo_2023_age_metallicity_for_disk_stars_in_MW_simulations} investigated intrinsic dispersion at fixed metallicity for all disk stars in m12i, and found it to be $\approx$ 0.025 dex at [Fe/H] = -0.25 dex and $\approx$ 0.015 dex at [Fe/H] = 0 dex.
The dispersion of intra-cluster dispersion shown by the vertical indigo splines spanning $\pm 1 \sigma$ of the intra-cluster dispersion tells us the width of the distribution of intrinsic scatters. These measurements are also very small which is an indication that our simulated clusters have some element scatter that is non-zero which is similar for all clusters.

Similarly as indicated by the upper red curves, the inter-cluster dispersion measurements for all three simulations m12i, m12f, and m12m is larger than the intra-cluster dispersion as expected. We measure mean inter-cluster dispersion of $\approx$ 0.03 dex for m12i, $\approx$ 0.09 dex for m12f, and $\approx$ 0.02 dex for m12m. We further test if we can use individual element abundances to discriminate between clusters. To do this, we examine pairs of stars within and between clusters to measure their chemical similarity. A requirement of reconstructing the formation conditions of  stars using only their individual abundances is that the chemical distance between stars within a cluster (intra-cluster pairs) is smaller than that of stars between clusters (inter-cluster pairs).  A minimum requirement that a pair of stars are from the same cluster is that they have the same [Fe/H]. We therefore now work only in a narrow [Fe/H] bin near the overall mean [Fe/H] of the OCs in each simulation.

\subsection{Intra and Inter-Cluster Chemical Difference Metric}
\label{3.3 chem diff metric analysis}

In this section, we discuss the chemical difference metric to see if we can distinguish between two stars that are formed in one OC and two stars randomly drawn from different OCs.
We employ the chemical difference metric described in \S \ref{subsec: chemical difference metric}, which calculates the squared chemical difference between two stars for each element and is summed over all elemental abundances. Such a metric allows us to consider how chemically similar two stars are via one number, which is small (approaching zero) when two stars are similar.

\begin{figure}  
    \includegraphics[width=\columnwidth]{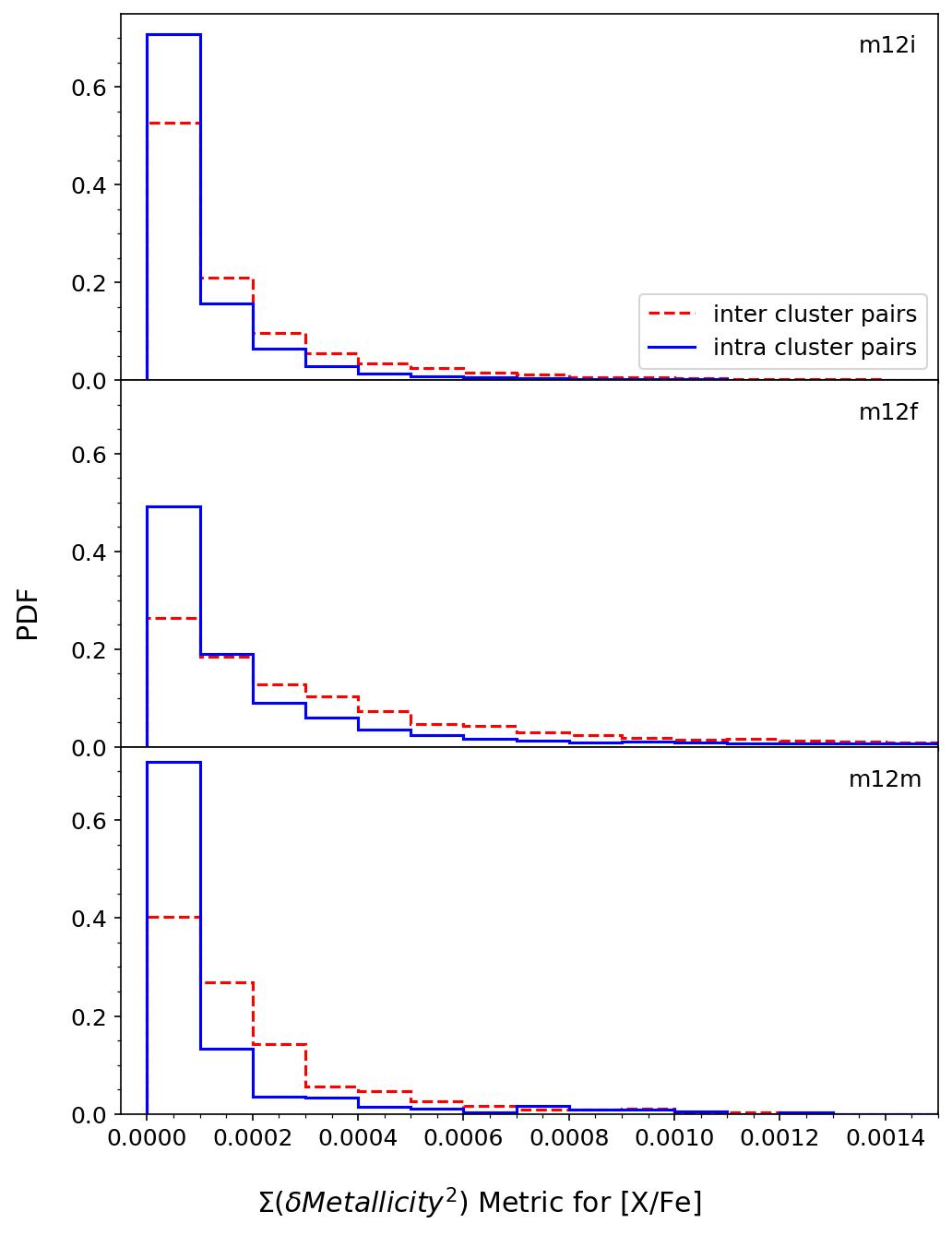}
    \caption{Histograms of the chemical difference metric for pairs of stars within OCs indicated by blue histograms(intra-cluster pairs) and between OCs indicated by the dashed red histograms(inter-cluster pairs) for each simulation m12i(top), m12f(middle) and m12m(bottom) in each panel. These clusters were selected to fall within $\pm{1\sigma}$ from the overall mean distribution of [Fe/H] in each simulation. The bin with the smallest chemical difference metric obtained for both inter and intra-cluster pairs is abundantly populated indicating that such difference is pretty small to uniquely identify the chemistry of individual clusters.}   
    \label{fig: chemical difference metric }
\end{figure}

We take the same sample of OCs highlighted in blue from Fig.~\ref{fig:feh vs R_CM} that is within $\pm 1 \sigma$ of the galaxy-wide mean  [Fe/H].
We use this sample to calculate the chemical difference between member stars to explore if we can discern the difference within and between clusters.
We first draw all possible pairs of stars from within an OC and calculate the chemical difference for each pair. This procedure is iteratively repeated for each OC in our sample and the results are presented in the blue histogram in Fig.~\ref{fig: chemical difference metric }. 
To maintain a comparative framework, we applied the same metric for an equivalent number of random pairs but this time, each star is drawn from a different randomly selected OC. 
This distribution is shown in the red histogram in Fig.~\ref{fig: chemical difference metric }. We form 17,747 intra and inter-cluster pairs in m12i, 1,685 pairs in m12f, and 374 in m12m.

In Fig.~\ref{fig: chemical difference metric }, the blue histogram consists of the difference metric for the pairs of stars drawn from the same OC. The majority of the pair's chemical difference falls within the smallest bins of the histogram for all three simulations.
If two stars are measured with such a small chemical difference, one would naively assume they come from the same OC.
However, these smallest chemical difference metric bins are also significantly populated by stars from inter-cluster pairs, as indicated by the red histograms. 
In m12i, $\approx$ 86 / 73\%  of the total intra / inter-cluster pairs have a difference metric of $<$ 0.0002. 
Similarly, in m12f, $\approx$ 66 / 43\%  of the total intra / inter-cluster pairs have a difference metric of $<$ 0.0002.
Finally, in m12m, $\approx$ 84 / 65\%  of the total intra / inter-cluster pairs have a difference metric of $<$ 0.0002.
Because of this, if we observe two stars with a chemical difference of $<$ 0.0002, it would be extremely difficult to know if these two stars came from the same OC or were drawn from two different OCs.
\footnote{We choose $<$ 0.0002 as a selection criteria here, as it is close to the mean chemical difference in all three simulations (see Table:\ref{Table:chemical_difference_metric_statistics} for relevant statistics for the chemical difference metric for each simulation. 
In Appendix~\ref{sec:Appendix_chemical_difference_metric_using_3_elements} we explore the impact of changing the statistic we use and of the number of elements we leverage to discern between intra and inter-cluster pairs.
Table:\ref{Table: Appendix chemical difference metric using all and few elements stats} shows that regardless of the statistic we use for subdividing our sample for the smallest chemical difference metric values, we always see significant contamination with inter-cluster pairs. 
We note that using fewer elements to calculate the chemical difference metric marginally decreases the ability to discern between the intra and inter-cluster pairs at small values of the chemical difference metric.}
That is, for the majority of pairs, it would be difficult to distinguish between stars that fall within the same OCs or different OCs using the vector of abundances for these elements alone.
Given that typical uncertainties in the abundance measurements are $\approx$ 0.03 dex \citep{ness_2018_doppelgangers}, it is unlikely that strong chemical tagging is feasible.
To further illustrate the impact of uncertainties in observed chemical differences, we calculate the chemical difference metrics for intra and inter-cluster pairs after adding the APOGEE errors, which we discuss in \S\ref{Sec: 3.4 Mertic with APOGEE Errors}.

To better quantify the degree of inherent similarity between the two distributions shown in Fig.~\ref{fig: chemical difference metric }, we perform a two-sample Kolmogorov-Smirnov (K-S) test.
Essentially, such a K-S test quantifies the likelihood that the observed datasets were drawn from the same underlying distribution and generates the probability that observed differences between the two in-hand datasets is due to chance selection. 
There are two metrics reported from a K-S test: the K-S statistic and the p-value; both metrics need to be considered together to fully understand the significance of the differences between the observed datasets.
The K-S statistic measures the maximum discrepancy between the cumulative distributions of two datasets.
The p-value indicates the probability of observing such a discrepancy by random chance alone if the null hypothesis is true; in a two sample K-S test, the null hypothesis is that the two observed datasets were drawn from the same underlying distribution.
Values of the K-S statistic and the p-value range between 0 and 1.
A high value of K-S statistic (> 0.6) and a small p-value (< 0.05\footnote{What is considered a ``small'' p-value is based on the chosen level of significance. 
In this work, we choose the level of significance to be 5\% thus we can trust the K-S statistic if the p-value is < 0.05.}) suggests that the observed samples are very different and are very likely to be drawn from two different underlying distributions.
Whereas, a low value of K-S statistic (< 0.4) and a small p-value (< 0.05) suggests that while the observed samples are similar, it is still very likely that they are drawn from two different underlying distributions.

We obtain a K-S statistic of 0.19 for m12i, which indicates a small maximum difference between the CDFs of the calculated chemical difference metric for the inter and intra-cluster pairs. 
However, the p-value is quite small ($\sim$$0.00$ for m12i), thus indicating that it is unlikely that these two samples are drawn from the same  global distribution.
In other words, while there are some similar aspects to both datasets (the maximum discrepancy to the global CDF is small), they are not in fact drawn from the same parent distribution.
This can be seen visually by inspecting the red and blue lines in the top panel of Fig.~\ref{fig: chemical difference metric }; while both distributions are strongly peaked at the lowest values of the chemical difference metric, at larger values, the shape of the inter-cluster distribution (shown in red) is different from the shape of the intra-cluster distribution (shown in blue).
The K-S statistic is 0.25 for m12f; this is slightly higher than m12i's K-S statistic, but still a critically low value, indicating that there is similarity in the two observed m12f datasets.  
However, again, the is p-value very low ($\sim$$0.00$ for m12f), which indicates that it is highly unlikely that both datasets are drawn from the same underlying distribution.
Finally, m12m has the largest K-S statistic ($\sim$$0.37$) of all three simulations; however, it still falls below the $0.4$ threshold, thus we can say that the maximum differences in m12m's distributions are quite slight.
Despite this, yet again, the p-value is vanishingly small ($\sim$$0.00$ for m12m), indicating that it is extremely unlikely that these samples were drawn from the same parent distribution.

Overall, we conclude while the intra and inter-cluster chemical difference metric datasets were drawn from two different global distributions, there is little power with this set of elements
in using small values of the chemical difference metric to select stars that definitively came from one OC.
That is, the contamination rate of inter-cluster pairs at the smallest chemical difference metric is far too significant to trust OC reconstruction.

\subsection{Intra and Inter-Cluster Chemical Difference Metric with APOGEE errors in simulated OCs}
\label{Sec: 3.4 Mertic with APOGEE Errors}

In order to test how the incorporation of errors in abundance measurements affects our chemical difference metric calculations, in this section, we add errors drawn from an observational cluster catalog to our simulated OCs. 
The errors we use come from OCCAM VI \citep[Open Cluster Chemical Abundances and Mapping Survey, ][]{natalie_myers_2022_OCCAM}, which was derived from APOGEE DR17 \citep{APOGEE_2022}.
To add errors to the simulation, we first generate a normal distribution for each element for each star; the mean of the distribution is set to the true value from the simulation.
We set the standard deviation of the distribution by calculating the mean error for each element from the OCCAM cluster catalog. 
We then draw from this distribution a single value to represent the chemical abundance with error included that is assigned to each star.
This process is repeated for all elements and all stars.
In this analysis, we use the 7 elements ([C/Fe], [N/Fe], [O/Fe], [Mg/Fe], [Si/Fe], [S/Fe] and [Ca/Fe]) that are in common with the OCCAM sample.
We use the same sample of clusters discussed in \S \ref{3.3 chem diff metric analysis} for our analysis here.

\begin{figure}
    \includegraphics[width=\columnwidth]{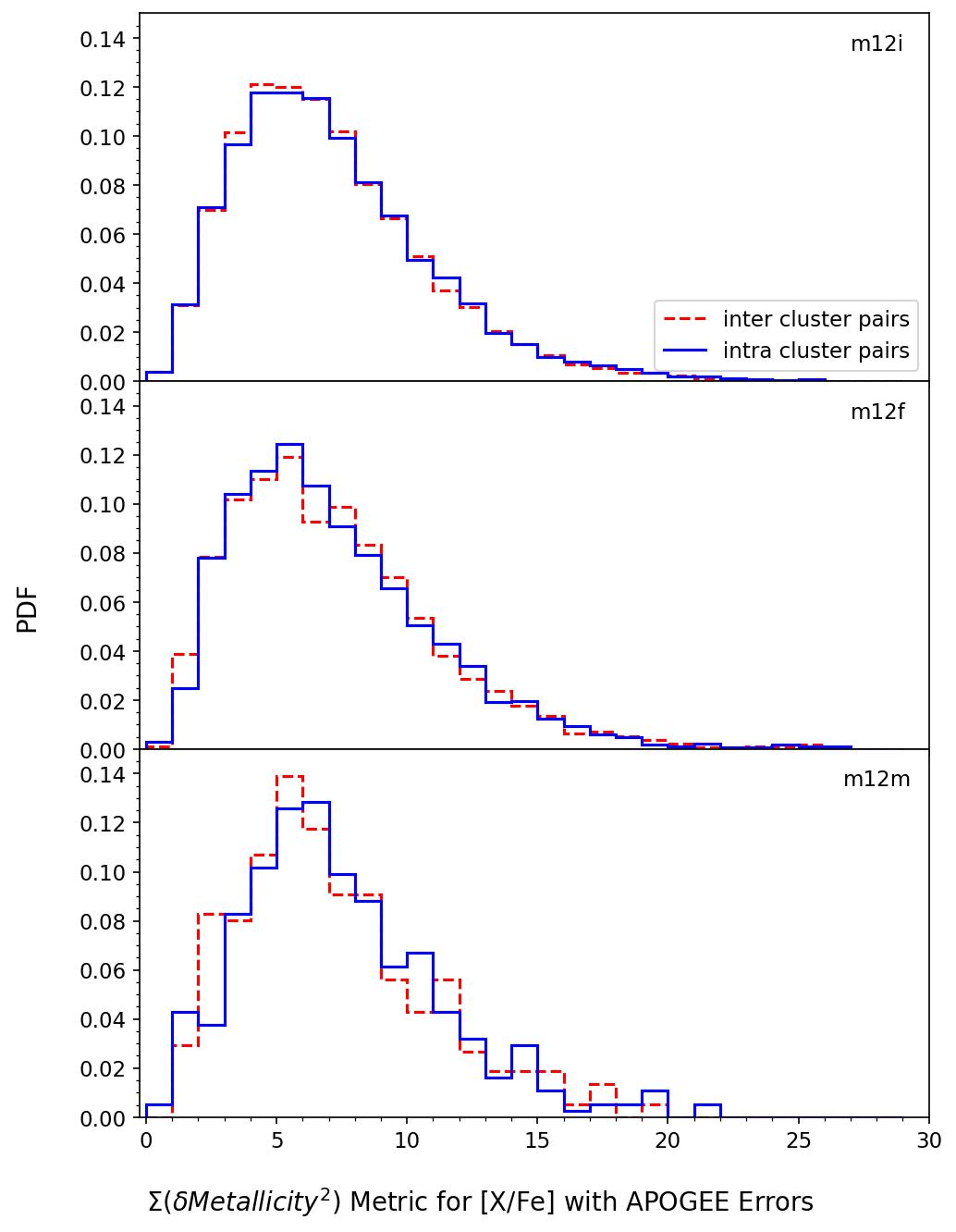}
    \caption{Histograms of the chemical difference metric for pairs of stars with mock APOGEE errors in OCs (blue) and between OCs (dashed red) for each simulation. These clusters were selected to fall within the mean $\pm{1\sigma}$ of the typical intra-cluster dispersion from the overall mean distribution of [Fe/H] in each simulation. The bins with the most of the chemical difference metric obtained for both inter and intra-cluster pairs are abundantly populated indicating that such difference is pretty small to uniquely identify the chemistry of individual clusters. }
    \label{fig: chemical difference metric APOGEE Errors }
\end{figure}

In Fig. \ref{fig: chemical difference metric APOGEE Errors }, each panel shows the histograms for the chemical difference metric from the intra and inter-cluster star pairs with APOGEE errors included; we show each distribution with a blue solid and red dashed line respectively. 
Again, the top, middle, and bottom panels correspond to m12i, m12f, and m12m respectively.
With errors included, the shape and the extent of the distribution shown in Fig. \ref{fig: chemical difference metric APOGEE Errors } changes substantially.
The distribution now peaks closer to the degree of freedom which is equal to the number of elements used to calculate the chemical difference metric. 
That is, each element contributes equally to the error-included chemical difference metric. 
Unlike histograms of the metrics with no error in abundance measurements that peaked at the smallest chemical difference, the histograms now peak at $\approx$ 7. All three panels from different simulations show a similar trend with the addition of errors and most of the bins until the metric value of 15 for each simulation are almost equally populated by both intra and inter-cluster pairs. 
Considering a limit of 7 which is closer to the mean of chemical difference metrics for both intra and inter-cluster pairs, $\approx$ 55\ / 56\%  of the total intra / inter-cluster pairs have the difference metric of $<$ 7 in m12i.
Similarly, for m12f, $\approx$ 55 / 54\% of the total intra / inter-cluster pairs have a difference metric of less than 7.
For m12m, the percentages of pairs with the difference metric of $<$ 7 are $\approx$ 52 / 56\% for intra / inter-cluster pairs respectively.
Also, the maximum PDF obtained for either the intra and inter-cluster pairs is slightly below $\approx$ 0.13 at the mean difference metric thus indicating there is no prospect of identifying if the star pairs are coming from the same cluster or different clusters. This analysis clearly shows that with errors, there is not enough chemical difference between the stars to identify them uniquely. 

\section{Discussion $\&$ Conclusions}\label{sec:discussion_conclusions}

In this paper, we have characterized the degree of scatter in abundance distributions in young OCs identified in three galaxies from the \emph{Latte} suite of FIRE-2 simulations (m12i, m12f and m12m).
Using a range of elements (C, N, O, Ne, Mg, Si, S, Ca, Fe) that are generated from three distinct nucleosynthetic families (core-collapse supernovae, white-dwarf supernovae, and stellar winds), we have shown that the abundance dispersion in FIRE OCs is comparable to observations from \citet{ness_2018_doppelgangers}. 

In this section, we compare our results with other observational studies that have measured the level of chemical homogeneity present in individual OCs using spectroscopic analysis.
Dispersions measured from observations are influenced by two factors: intrinsic scatter -- the true difference in elemental abundance between stars within an OC -- and observational noise, which originates from measurement uncertainties.
Most observational studies consider an OC to be chemically homogeneous if the metallicity dispersion inside a cluster is less than or equal to the typical measurement uncertainties.
While measurement uncertainties vary between studies
\citep[e.g., 0.03 dex vs.\ 0.008--0.036 dex, from][respectively]{ness_2018_doppelgangers, Liu_2016_inhomogeneity_hyades}, typically measurement uncertainty is $\leq$ 0.03 dex \citep{jo_bovy_2016_chemical_homogeneity_of_open_clusters,ting_2012_homogeneity_open_cluster_ic4756}. 

Simulations do not suffer from observational uncertainties, and the typical intrinsic dispersion inside a simulated \emph{Latte} OC is <~0.02 dex. This is broadly consistent with the intrinsic dispersion measured for these elements from observational studies.
In \emph{Latte}, such homogeneity is a consequence of two dominant factors: 1) at present day, gas within individual GMCs is very well-mixed \citep[see Figure 7,][]{bellardini_2021_3d_gas}, and 2) star formation within an individual OC is typically short-lived ($\sigma_{\text{age}} < 1$ Myr, Masoumi, et al.~in prep), which limits internal enrichment.

Note, some level of inhomogeneity is expected in observations driven by, among other things, atomic diffusion, mixing processes and planet engulfment \citep{blanco_cuaresma_2015_inhomogeneity_evolutionary_stages, bertell_motta_2018_inhomogeneity,spina_2021_planet_formation}.
Such processes are expected to have an impact on the metallicity of evolved populations and are not resolved in \emph{Latte} OCs, as star particles have a fixed abundance throughout their lifetime.
In observational studies, dispersion measurements that make use of both main sequence and giant stars within an OC are expected to have an increased metallicity dispersion; as a result, some studies consider dwarf and giant stars separately \citep{blanco_cuaresma_2015_inhomogeneity_evolutionary_stages, casamiquela_2020_inhomogeneity}.
Moreover, different types of elements (e.g., light versus heavy elements) are expected to have a different amount of scatter \citep{de_silva_abundance_patterns_2009,manea_2023_distinguishing_power_neutron_capture}.
While the current generation of \emph{Latte} simulations do not track heavy elements, such an improvement is anticipated in upcoming generations of FIRE.

With this somewhat complicated observational context in mind, we now present results from the literature.
Note, as much as possible, we try to include both the stellar populations and the elements studied in each work.
In one of the earliest spectroscopic studies on intra-cluster homogeneity, \citet{de_silva_2006_chemical_homogeneity_hyades} analyzed the heavy elements Zr, Ba, La, Ce, and Nd in dwarf stars in the Hyades OC; they showed that the chemical scatter within this system is below their measurement uncertainties, which ranged from 0.047--0.048 for Zr and Ba, to 0.025--0.026 dex for La, Ce and Nd. 22
Thus, \citet{de_silva_2006_chemical_homogeneity_hyades} conclude that Hyades is chemically homogeneous.
In a follow-up study, \citet{d_silva_2007_chemical_homogeneity} analyzed abundances of the light elements Na, Mg, Si, Ca, Mn, Fe, and Ni in 12 red giant stars in the OC Collinder 261, and estimated the intrinsic scatter is < 0.05 dex across these elements. 
Because their measurement uncertainty ranged between 0.01--0.05 dex, they concluded Collinder 261 is chemically homogeneous. 
They also compared Collinder 261, Hyades, and the moving group HR 1614 to each other and found all three systems have small scatter ($<$ 0.05 dex) consistent with homogeneity \citep{d_silva_2007_chemical_homogeneity}. 

In line with \citeauthor{de_silva_2006_chemical_homogeneity_hyades}, many other groups have found that OCs are chemically homogeneous from spectroscopic analysis.
For example, \citet{pancino_2010_3_red_clump_stars_5_OCs_homogeneity} analyzed three red clump stars in each of five OCs: Cr110, M37, NGC 2420, NGC 7789, and M67, and they found the dispersion in [Fe/H] indicated chemical homogeneity for all five systems.
Likewise, using 12 subgiants, \citet{ting_2012_homogeneity_open_cluster_ic4756} showed that variation in [X/H] for Na, Al, Mg, Si, Ca, Ti, Cr, Ni, Fe, Zn and Ba is <0.03 dex in IC 4756. 
Additionally, \citet{cunha_2015_na_O_in_NGC6791} studied Na and O abundances for 11 APOGEE red giants in the old, super-solar OC NGC 6791, with measurement errors of $\sim$0.05 dex. 
They concluded this system is chemically homogeneous, with an abundance dispersion of $\sim$0.05--0.07 dex; this conclusion is consistent with results for NGC 6791 presented by \citet{bragaglia_homogeneity_2014}.
\citet{jo_bovy_2016_chemical_homogeneity_of_open_clusters} also used APOGEE spectra for 49 giants in M67, NGC 6819, and NGC 2420 and showed that the scatter is <0.03 dex for 15 elemental abundances (C, N, O, Na, Mg, Al, Si, S, K, Ca, Ti, V, Mn, Fe, and Ni). 
Among these elements, C and Fe had strong limits ($<$ 0.02 dex at 95$\%$ confidence); however, for Na, S, K, Ti, and V, the scatter was <0.05 dex.
\citet{poovelil_2020_oc_chemical_homogeneity_MW} also calculated the abundance scatter in APOGEE red giants in 10 OCs for 8 elements, Mg, Al, Si, Ca, Fe, Si, Mn, and Ni, and found that the dispersion for these clusters were within the limits  
 indicated by \citet{jo_bovy_2016_chemical_homogeneity_of_open_clusters} of <0.03 dex for most of the elements, with only Mg, Al, and Si having larger scatter, which could be consistent with dredge-up effects that preferentially impact light elements. 
 
 However, some studies have found measurable chemical inhomogeneity larger than the measurement uncertainties in OCs \citep[e.g.,][]{blanco_cuaresma_2015_inhomogeneity_evolutionary_stages,Liu_2016_inhomogeneity_hyades,bertell_motta_2018_inhomogeneity,casamiquela_2020_inhomogeneity}.
 For example, \citet{Liu_2016_inhomogeneity_hyades} studied the abundances of 19 elements (C, O, Na, Mg, Al, Si, S, Ca, Sc, Ti, V, Cr, Mn, Fe, Co, Ni, Cu, Zn, and Ba) in 16 solar-type stars in the Hyades cluster, with abundance errors ranging from a low of 0.008 dex for Si to a high of 0.036 dex for S.
 For most elements, the abundance dispersion they measured was significantly larger than measurement errors, by a factor of $\approx$1.5 - 2.  
 They conclude that Hyades is chemically inhomogeneous at $\ge$0.02 dex level, which contrasts with the conclusions for Hyades drawn from \citet{de_silva_2006_chemical_homogeneity_hyades} and \citet{d_silva_2007_chemical_homogeneity}.
 \citet{Liu_2016_inhomogeneity_hyades} suggest pollution of metal-poor gas into portions of the proto-cluster cloud combined with supernova ejection into other regions with incomplete mixing is the root cause of such dispersion.
 
Overall, we find that different observational studies have differing criteria to determine if an OC is chemically homogeneous. 
This conclusion can be impacted by the stellar sample used (containing evolved, main sequence or a combination of the two populations), elements considered (light and/or heavy elements), and measurement errors (which is instrument resolution and line-fitting methodology dependent).
Dwarf stars and heavy elements show the greatest discriminating power for the measurement of homogeneity. 
Most studies agree that the typical level of intra-cluster scatter inside Milky Way OCs is $\leq$ 0.03 dex for most elements; OCs with scatter less than this per element relative to H are considered chemically homogeneous.
In our simulations, the intrinsic OC intra-cluster scatter in [X/H] for C, N, O, Ne, Mg, Si, S, Ca, Fe is $\leq$ 0.02 dex; these \emph{Latte} OCs would be considered chemically homogeneous by most observational studies.

While there is abundant research supporting small intra-cluster dispersion in OCs, the assumption of large inter-cluster dispersion, which is crucial for strong chemical tagging to work, has only recently been explored \citep[e.g.,][]{blanco_cuaresma_2015_inhomogeneity_evolutionary_stages, lambert_reddy_2016_inter_cluster_dispersion}. 
The first study to systematically consider inter-cluster dispersion was \citet{de_silva_abundance_patterns_2009}. 
They aggregated abundance information from the literature for 10 elements (Na, Si, Ca, Mn, Zn, Rb, Zr, Ba, Nd, and Eu) from 24 OCs and showed that different clusters exhibit varying levels of dispersion for each element.
Despite likely systematic differences in their sample due to varying methodologies and scales, as well as the inclusion of both evolved and main sequence stars introducing additional scatter in inter-cluster dispersion (particularly in light elements like Na), they were still able to conclude that different clusters exhibit differing elemental abundance patterns.
In particular, they found an inter-cluster scatter of $\geq$0.20 dex in heavy elements like Mn and Ba, which they concluded is an indication of abundance variation between clusters.  

In the time since, there have been several significant studies that investigate
chemical abundances in large spectroscopic samples of OCs surveyed in homogeneous manner. 
These include the Bologna Open Clusters Chemical Evolution project \citep[BOCCE][]{BOCCE2006}, the WIYN Open Cluster Study \citep{WIYN2000, WIYN2011}, Gaia ESO OC papers such as \citet{Magrini2014}, and APOGEE  studies such as the Open Cluster Chemical Abundance and Mapping Survey  \citep[OCCAM][]{natalie_myers_2022_OCCAM}.
While these studies provide tremendous insight into radial trends in the Milky Way, unfortunately, as of yet, they do not explicitly consider inter-cluster trends.

Subsequent to \citet{de_silva_abundance_patterns_2009}, \citet{mitschang_2013_distance_metric} used 30 distinct OCs from the literature and measured both the inter and intra-cluster dispersion.
They found a typical inter-cluster dispersion of $\sim$0.120 dex and considerably smaller intra-cluster scatter of $\sim$0.045 dex (i.e., the inter-cluster dispersion is $\sim$3 times greater than the intra-cluster dispersion).
Thereafter, the first study (to our knowledge) that considered inter-cluster scatter is \citet{blanco_cuaresma_2015_inhomogeneity_evolutionary_stages}.
In this work, they compiled stellar spectra from 31 old and intermediate-age OCs to determine 17 abundance species and showed that stars at different evolutionary stages have distinct chemical patterns. Upon separating stars into dwarfs and giants, they found only a few OCs show distinct chemical signatures, with the majority of them possessing a high degree of overlap.
They also analyzed five OCs (IC4651, M67, NGC2447, NGC2632, and NGC3680) with stars in various evolutionary stages and
found that differences in abundances for [Fe/H] and [X/Fe] were less than 0.05 dex for almost every element considered, except for Na, which can be enhanced in giants. 
Such enhancement likely occurs as a result of mixing from the first dredge-up in giant stars. 

On the  other hand, \citet{lambert_reddy_2016_inter_cluster_dispersion} studied inter-cluster abundance differences for 23 elements in 70 red giants from 28 OCs. 
Their analysis was comparable in scope and accuracy to that reported by \citet{blanco_cuaresma_2015_inhomogeneity_evolutionary_stages}, except that their selection included several additional heavy
elements -- La, Ce, Nd, Sm, and Eu -- which are produced by neutron-capture processes. Using a set of 20 clusters, they calculated scatter in light elements which was 0.16 dex for Na and Al,  0.12 dex for Mg and Si and 0.11 dex for Ca \citep[see Table 1 from][]{lambert_reddy_2016_inter_cluster_dispersion}.
They observed inter-cluster scatter up to 0.5 dex for heavy elements like La and Ce, with uncertainty <0.07--0.08 dex.
\citet{lambert_reddy_2016_inter_cluster_dispersion} point out that while the inter-cluster scatter for heavy elements is large, many clusters are individually chemically homogeneous in heavy elements with unique chemical patterns.
They suggest that combining measurements of one or more light elements like Mg, Al, Si, Cu, and Zn with heavy elements like La, Ce, Nd, Sm, and Eu in future chemical tagging studies might produce insightful results \citep{lambert_reddy_2016_inter_cluster_dispersion}. 

Overall, at present, there is limited discussion of inter-cluster scatter in the literature and differing perspectives offered in the two most recent works.
\citet{blanco_cuaresma_2015_inhomogeneity_evolutionary_stages} have found a high degree of overlap (small inter-cluster scatter) in the chemical signatures of light elements across OCs; however, they don't directly quantify the inter-cluster scatter in their analysis. 
Whereas, \citet{lambert_reddy_2016_inter_cluster_dispersion} have found a significant inter-cluster scatter in heavy neutron-capture elements (up to 0.5 dex) and a smaller scatter in ligher elements (on average $\approx$0.13 dex).
As previously mentioned, at this time, we only track light elements in our simulations.
In this limit, the OC inter-cluster
scatter we measure for [X/H] for C, N, O, Ne, Mg, Si, S, Ca, $\&$ Fe is 2$-$5 times larger than the intra-cluster dispersion in individual OCs. 
Specifically, when averaged across all 9 elements, the inter vs intra-cluster dispersion is for m12i: 0.03 vs 0.01, for m12f: 0.09 vs 0.02 and for m12m: 0.02 vs 0.01. 
We find that when inter-cluster scatter is considered by element, only C and N have a slightly higher scatter than the average (albeit it at a very small level, see Fig. \ref{fig:5 intra and inter cluster dispersion}). This is consistent with the result presented by \citet{mitschang_2013_distance_metric}, which showed that the inter-cluster dispersion is $\sim$3 times
greater than the intra-cluster dispersion.
\citet{lambert_reddy_2016_inter_cluster_dispersion} have shown inter-cluster dispersion to be >10 times larger than the intra-cluster dispersion for heavy elements like La, Ce, Nd and Sm (e.g., 0.4 dex of intra-cluster scatter vs 0.03 dex of inter-cluster scatter for La). 
However, if we assume that the individual OCs from \citet{lambert_reddy_2016_inter_cluster_dispersion} are homogeneous at $\approx$0.03 dex level, then the ratio of inter to intra-cluster scatter for light elements is $\sim$4, which is in line with our results.

As one might expect, the inter-cluster scatter that we report for OCs is less than what has been seen for \emph{all} young stars in the \emph{Latte} disks.
For  example,
\citet{matthew_bellardini_2022_3d_abundances_FIRE} selected young stars ($<$ 500 Myr old) at present day from 11 of the \emph{Latte} galaxies at a range of radii and showed that at small radii, the scatter across the entire annulus is 0.03--0.04 dex.
Moreover, at present day, the typical variation across the fiducial solar cylinder (R = 8 kpc) in [Fe/H] is $\approx$ 0.05 dex \citep[see top panel Figure 9,][]{matthew_bellardini_2022_3d_abundances_FIRE,graf_2024_spatial_variation_abundances_current_and_past}.
However, at this annulus, the inter-cluster scatter for OCs in [Fe/H] is only 0.02 dex.

In addition to intra and inter-cluster scatter, we have calculated a chemical difference metric similar to the one first introduced by \citet{mitschang_2013_distance_metric}.
Such a metric helps to determine if one can distinguish pairs of stars formed in the same OC from pairs formed across different respective OCs.
We calculated the percentage of pairs that have small chemical difference metric values ($<$ than the mean chemical difference for the entire sample) with the hope that this population would be predominantly populated by intra-cluster pairs. 
However, for m12f, 43\% of the inter-cluster and 66\% of the intra-cluster pairs had similar indistinguishably small chemical difference values. 
Moreover, in the other two galaxies, the rate of chemical difference overlap was even greater:  65\% vs 84\% for m12m and 73\% vs 86\% for m12i in inter vs intra-cluster pairs.
We find it noteworthy that a significant fraction of the smallest chemical difference values belong to inter-cluster pairs, which one would typically assume to be populated by only intra-cluster pairs.

\citet{mitschang_2013_distance_metric} used the chemical difference metric to compare 30 distinct OC from across the literature. 
Their abundance measurements ranged from 5 to 23 elements per cluster across 291 stars. 
They found a distinct peak in the distribution of the chemical difference metric for both the intra and the inter-cluster distributions.  
In their analysis, the two distributions had a critical crossing point at a chemical metric value of $\sim$0.07 --- at this value, both the intra and inter-cluster star pairs showed identical chemical differences distributions and thus the metric had no constraining power for strong chemical tagging.
In our simulated OCs, such a crossing-point occurs at a very small chemical difference value ($\sim$0.0001); this crossing point is near the median of both skewed Gaussian distributions, and is very well sampled by both intra and inter-cluster pairs. 
Notably, the chemical difference metric has no constraining power to determine origin at this location.
In our extended analysis of the chemical difference metric, we added APOGEE errors to the abundance measurements of our OCs. 
We find that the error added chemical metric distributions for intra and inter-cluster pairs overlap significantly at almost all measured values, further reinforcing that there is little prospect of discerning intra and inter-cluster pairing using the chemical difference metric from light elements alone.

In a slightly different scenario, leveraging 600 intra-cluster pairs and 1,018,581 field pairs with similar $\text{log}(\text{g})$ and $T_{\text{eff}}$ values, \citet{ness_2018_doppelgangers} utilized 20 element abundances to calculate the chemical difference metric. 
The chemical difference distribution for intra-cluster and field star pairs were visibly distinct but not totally disjoint, with 0.3$\%$ of field stars having similar chemical differences as the median  difference in intra-cluster pairs. 
When stars of a fixed solar metallicity ([Fe/H]=0$\pm$0.02) were considered, the percentage of field star pairs with overlapping intra-cluster chemical difference values increased to $\sim$1.0$\%$.
\citet{ness_2018_doppelgangers} refer to these field stars as solar "doppelgangers." 
They conclude that such a significantly populated doppelganger population implies that strong chemical tagging in a strict sense would not work with this dataset.
While we are intrigued by this result, we leave a similar calculation utilizing our OC sample and \emph{Latte} field stars to future work. 

In conclusion, OCs in FIRE show realistic intra-cluster homogeneity (typical OC elemental dispersion $\lesssim$0.02 dex) and comparable inter-cluster dispersion to observations (2-5 times larger than intra-cluster dispersion).
However, when using a chemical difference metric to distinguish between intra and inter-cluster pairs, we do not find significantly large chemical differences to distinguish OC origin for the majority of stars.
This is likely due to limited and/or redundant information encoded in the light elements we track (C, N, O, Ne, Mg, Si, S, Ca, Fe); with the addition of heavy elements, the chemical difference metric may produce more reliable results for chemical tagging in the future.
However, as informed by our simulated OC sample (drawn from three realistic Milky Way-mass cosmological galaxy simulations with varied merger histories and environments), in the  limits of light elements only, we are dubious that strong chemical tagging can reconstruct individual OCs reliably to high level of confidence.

\section*{Acknowledgements}
BB and SL would like to thank Peter Frinchaboy and Natalie Myers for their help with integrating APOGEE OCCAM errors into the simulation data.
BB acknowledges support from the Center for Computational Astrophysics (CCA), which enabled his remote participation in the 2021 pre-doctoral program.

BB acknowledges support from NSF grant AST-2109234. 
SL acknowledges support from NSF grant AST-2109234 and HST grant AR-16624 from STScI.
AW received support from: NSF via CAREER award AST-2045928 and grant AST-2107772; NASA ATP grant 80NSSC20K0513; HST grant GO-16273 from STScI.
ECC acknowledges support for this work provided by NASA through the NASA Hubble Fellowship Program grant HST-HF2-51502 awarded by the Space Telescope Science Institute, which is operated by the Association of Universities for Research in Astronomy, Inc., for NASA, under contract NAS5-26555.

We generated simulations using: XSEDE, supported by NSF grant ACI-1548562; Blue Waters, supported by the NSF; Frontera allocations AST21010 and AST20016, supported by the NSF and TACC; Pleiades, via the NASA HEC program through the NAS Division at Ames Research Center.

FIRE-2 simulations are publicly available \citep{wetzel_2023_FIRE2_data_release} at \url{http://flathub.flatironinstitute.org/fire}.
Additional FIRE simulation data is available at \url{https://fire.northwestern.edu/data}.
A public version of the \textsc{Gizmo} code is available at \url{http://www.tapir.caltech.edu/~phopkins/Site/GIZMO.html}.

\vspace{2 mm}
\noindent{\emph{Software:} IPython \citep{ipython}, Matplotlib \citep{matplotlib}, Numpy \citep{numpy}, Scipy \citep{scipy}, \texttt{halo\_analysis} \citep{haloanalysis}, \texttt{gizmo\_analysis} \citep{gizmoanalysis}.}

\bibliographystyle{aasjournal}
\bibliography{main}{}

\appendix
\vspace{-4mm}
\renewcommand{\thesection}{\Alph{section}}
\renewcommand{\thesubsection}{\thesection.\arabic{subsection}}


\begin{center}
\section{Correlation between different elements}\label{sec:Appendix_correlation_between_elements}
\end{center}

In this appendix, we consider the scatter and correlation between elemental abundances for all stars in all OCs found in m12i and presented in this paper.
A discussion of how yields are implemented in FIRE-2 can be found in Appendix A in \citet{andreia_carrillo_2023_age_metallicity_for_disk_stars_in_MW_simulations} and in \citet{hopkins_2018_fire2}.

\begin{figure*}
    \centering 
    \includegraphics[width=0.6\textwidth]{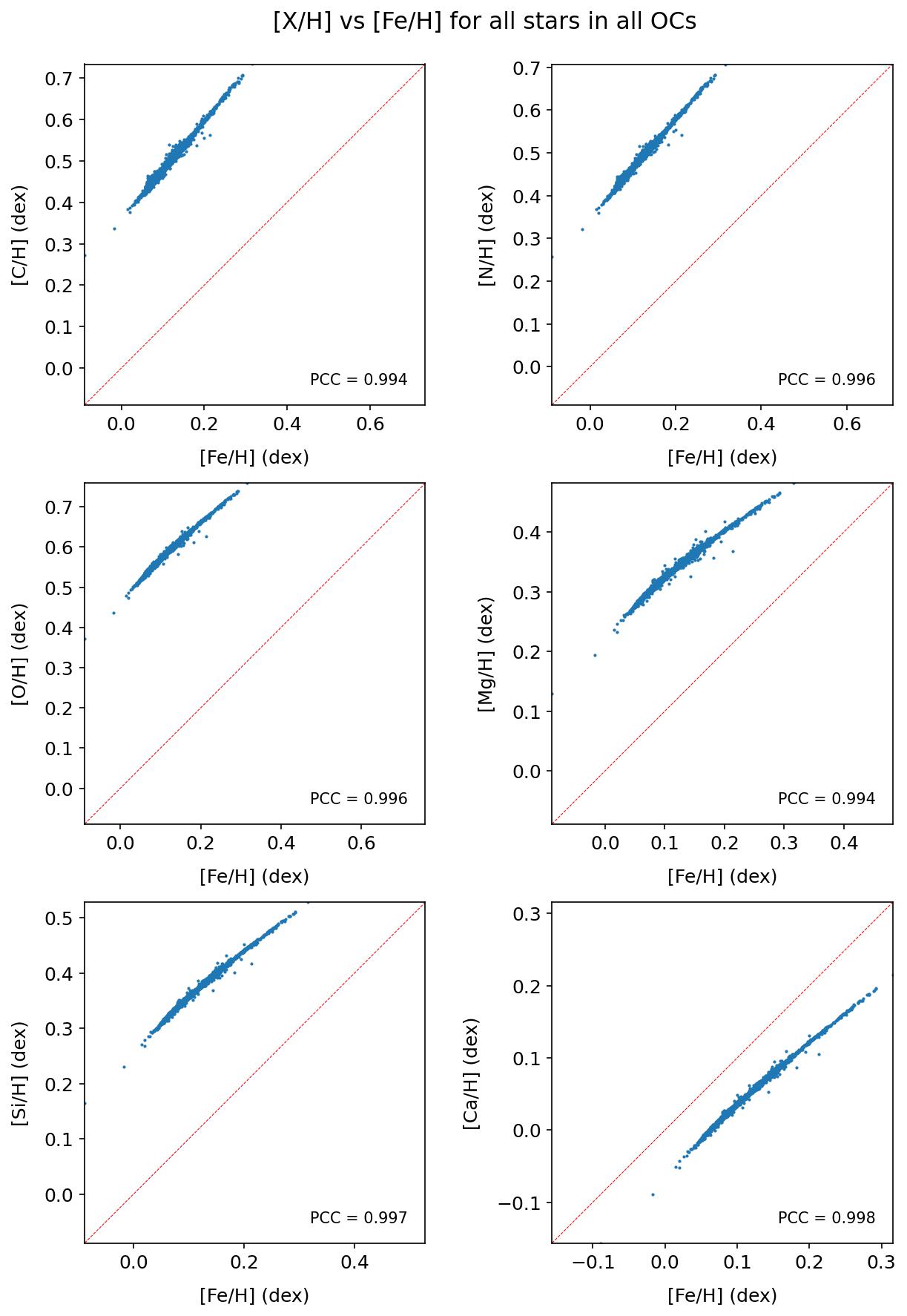}
    \caption{ Left to right, top to bottom: [X/H] vs [Fe/H] for C, N, O, Mg, Si, and Ca for stars from OCs presented in this paper in m12i. Elements are highly correlated with [Fe/H] as indicated by the Pearson correlation coefficient, which is labelled on each plot ($>$ 0.99 for all elements considered). The dotted diagonal red line is the one-one line. Note that the slope of the [X/H] vs [Fe/H] is similar to the one-one line and there is limited scatter in [X/H] at any value of [Fe/H]. This indicates that all the elements considered carry similar discriminating power as [Fe/H].}
    \label{fig: correlation between element over feh}
\end{figure*}

\begin{figure*}
    \centering 
    \includegraphics[width=0.95\textwidth]{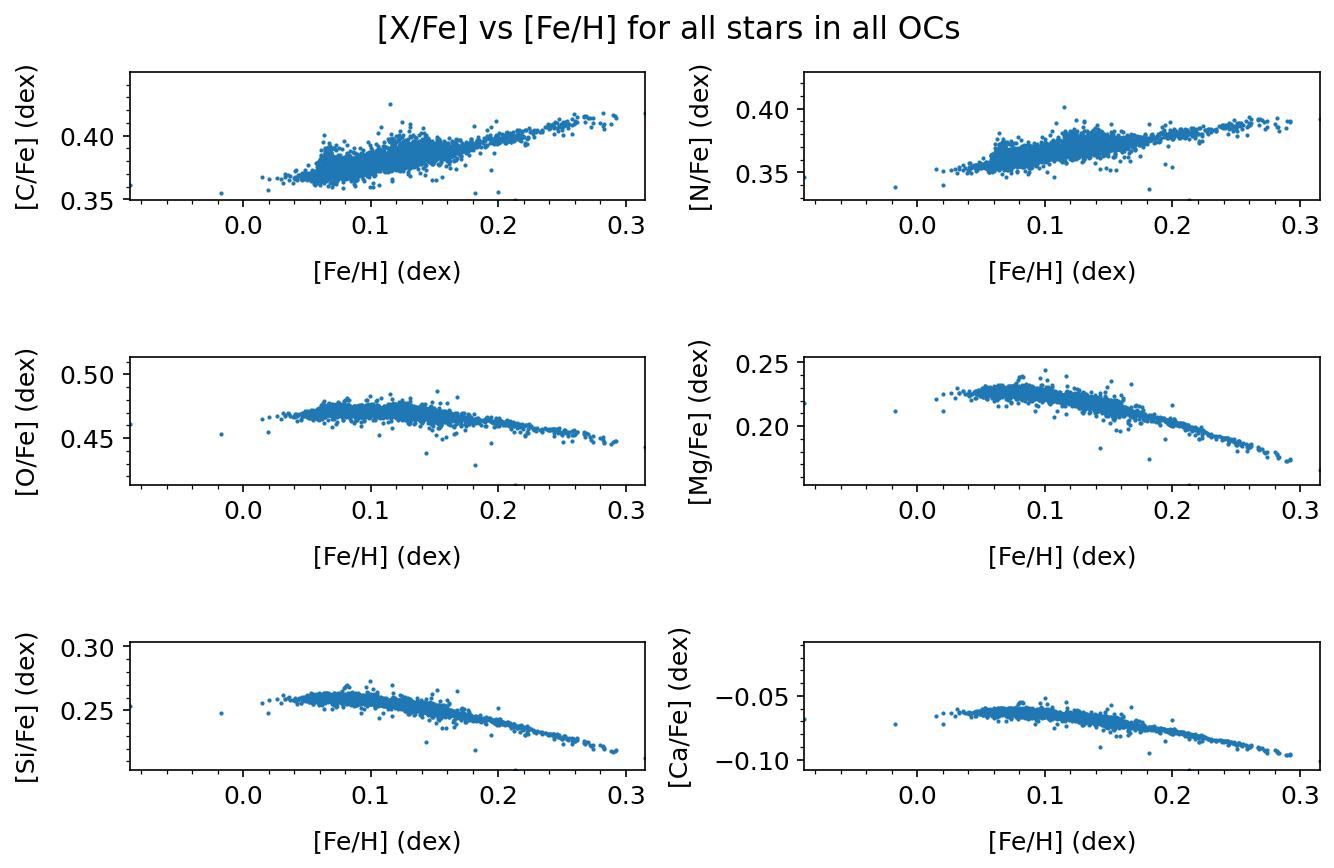}
    \caption{ Left to right, top to bottom: [X/Fe] vs [Fe/H] for C, N, O, Mg, Si, and Ca for stars from OCs presented in this paper in m12i. 
    Elements have a very tight scatter for a given value of [Fe/H], as indicated by the small range in [X/Fe] shown in y-axes. 
    The scatter is more substantial at lower values of [Fe/H]; however, the scatter decreases substantially at [Fe/H]$\ge$0.19 dex.} 
    \label{fig: scatter on xfe vs feh}
\end{figure*}

In Fig.~\ref{fig: correlation between element over feh}, we consider [X/H] versus [Fe/H] for X=C, N, O, Mg, Si, Ca. 
We find that each of these elements are highly correlated with [Fe/H], as indicated by the Pearson correlation coefficient ($>$ 0.99 in all cases). 
This can be seen by comparing the trend in the data in each panel to the diagonal red dotted one-to-one line. 
Not only is there limited scatter in elemental abundance at any value of [Fe/H] (as shown in Fig. \ref{fig: scatter on xfe vs feh}, which presents [X/Fe] versus [Fe/H] to highlight the limited scatter at any value of [Fe/H]), but also, the data in Fig.~\ref{fig: correlation between element over feh} follows a simple linearly increasing trend in all panels, with a slope that is nearly identical to the one-to-one line.
While there are some slight departures from a single linear distribution (e.g., at [Fe/H]$\sim$0.15, a transition to  a shallower slope for [Mg/H] and a transition to a steeper slope [C/H] and [N/H]), to first order, all the elements we consider in this paper carry largely the same chemical signature as [Fe/H].
That is, modulo a normative offset, given a value of [Fe/H] for any star in any of our OCs in m12i, [X/H] can be predicted with high confidence for C, N, O, Mg, Si, Ca.

\begin{figure*}
    \centering
    \includegraphics[width=0.75\textwidth]{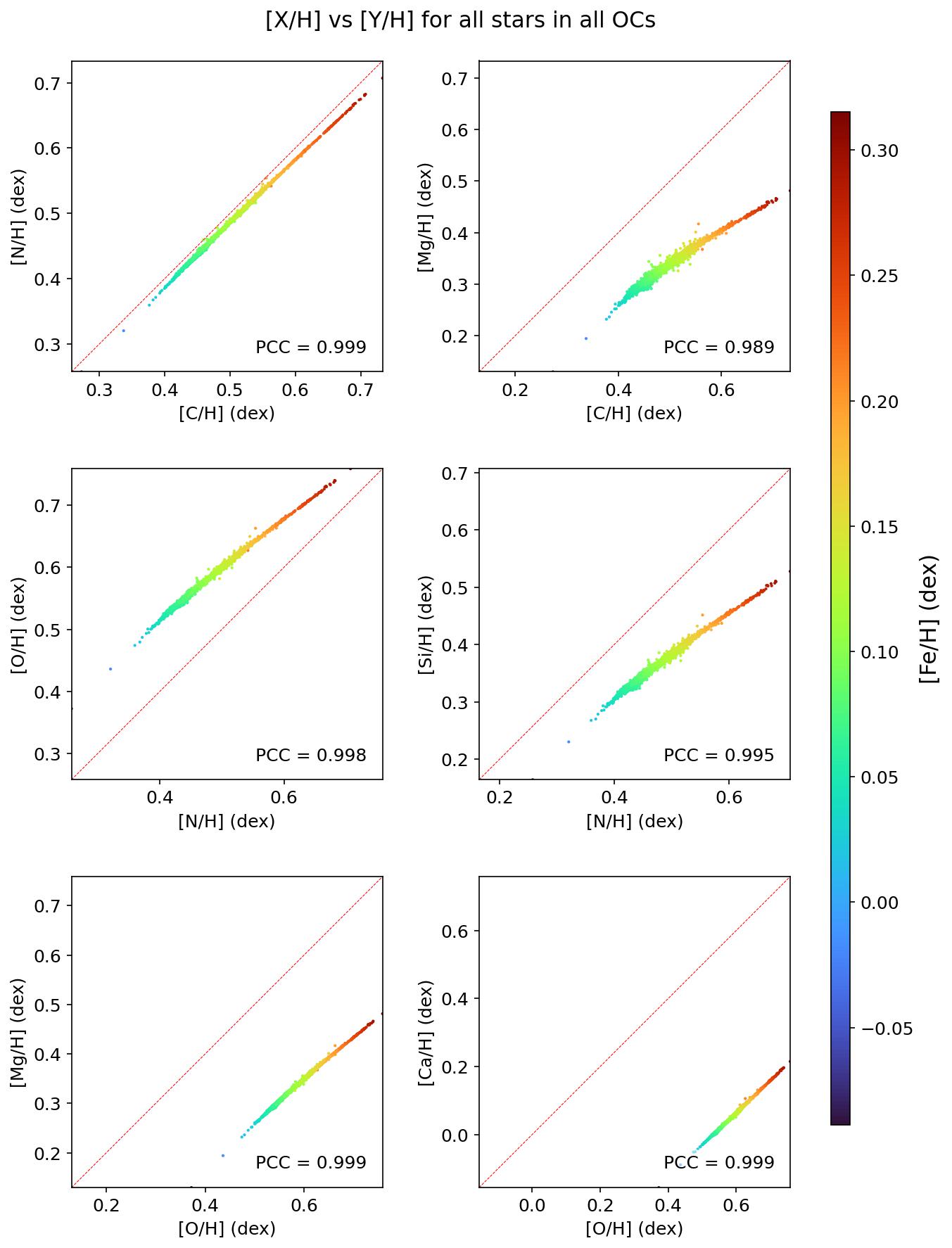}
    \caption{Correlation between [X/H] vs [Y/H] for C, N, O, Mg, and Ca color-coded by [Fe/H] (shown in the colorbar). Elements are highly correlated with one another, as indicated by the Pearson correlation coefficient ($\ge$ 0.989 for all elements considered). The dotted diagonal red line is the one-one line. Note that the slope of the [X/H] vs [Y/H] is similar to the one-one line and there is limited scatter in [X/H] at any value of [Y/H]. This indicates that all the elements considered carry similar discriminating power.}
    \label{fig: correlation between element over another element}
\end{figure*}

To further reiterate this point, in Fig.~\ref{fig: correlation between element over another element} we explore the relationship between individual elements with one another by plotting [X/H] versus [Y/H] for X,Y=C, N, O, Si, Mg, Ca. 
Here, we color code each star by its [Fe/H] value to reflect the trend shown in Fig.~\ref{fig: correlation between element over feh}.
Again, we see that these elements are highly correlated with one another when considering a Pearson correlation coefficient or inspecting the trends relative to the one-to-one line. 
Of all the elements, [Mg/H] is the least correlated with [C/H], with a Pearson correlation coefficient of 0.989.
Indeed, there is a slight scatter in [Mg/H] for small values of [C/H] and a modest transition in slope at high values of [C/H].
To this end, in our subsequent chemical difference metric analysis presented in Appendix~\ref{sec:Appendix_chemical_difference_metric_using_3_elements}, we utilize Mg and C as maximally independent elements. 
However, it bears repeating that all of the elements presented here carry very similar discriminating power. 
Note, while this analysis indicates that chemical information is redundant in our simulations, something similar has been found in the Milky Way. 
For example, \citet{griffith_2024_KPM} and \citet{ness_2022_homogeneity_of_star_forming_environment_in_the_milky_way} showed that if Fe and Mg are measured alone, then eight other supernova elements can be predicted to within 5$\%$ of their true values using a simple linear regression \citep[see Figure 1][for an analogous figure to Fig.~\ref{fig: correlation between element over another element}, but for the Milky Way]{ness_2022_homogeneity_of_star_forming_environment_in_the_milky_way}.

\vspace{4mm}
\begin{center}
\section{Chemical Difference Metric using only 3 elements}\label{sec:Appendix_chemical_difference_metric_using_3_elements}
\end{center}

Here we consider how the chemical difference metric is impacted by the number of elements used or the types of elements chosen to compute the metric. 
We have shown our analysis using all elements (C, N, O, Mg, Si, S, Ca, Ne) in \S\ref{3.3 chem diff metric analysis}. 
However, as shown by Figs. \ref{fig: correlation between element over feh}, \ref{fig: scatter on xfe vs feh}, and \ref{fig: correlation between element over another element}, elements are highly correlated with each other in our simulations. 
Thus, here we discuss the distribution of chemical differences in two other cases: calculated using only elements Mg, C and N, as well as calculated using only elements Mg, C and O. 
We have selected these elements as the maximally uncorrelated elements presented in Appendix~\ref{sec:Appendix_correlation_between_elements}. 
We compare these results alongside the case where we use all elements to compute the metric. 

\begin{figure*}
    \includegraphics[width=0.9\textwidth]{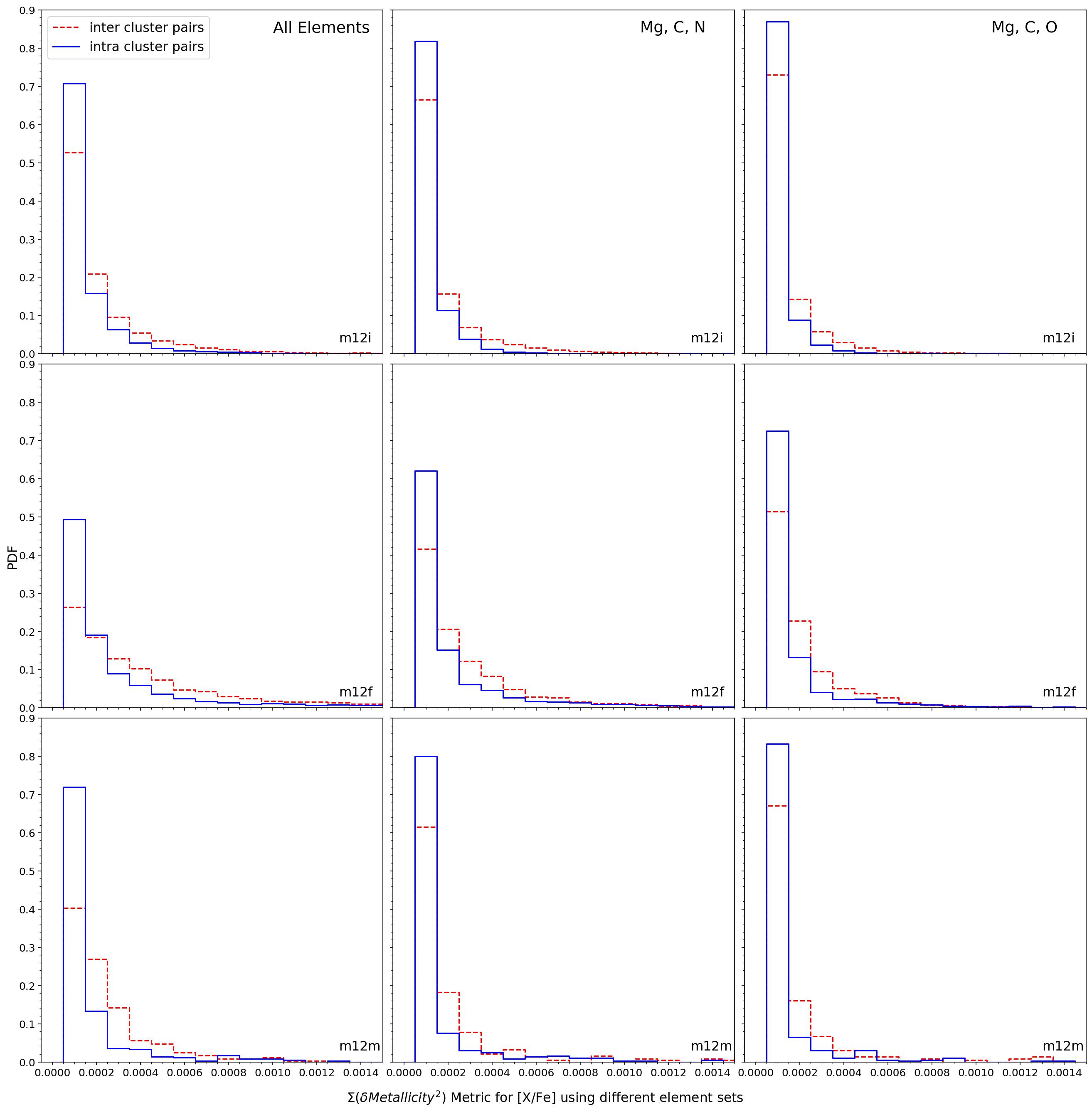}
    \caption{Histograms of the chemical difference metric for pairs of stars within OCs indicated by blue histograms (intra-cluster pairs) and between OCs indicated by the dashed red histograms (inter-cluster pairs) for each simulation m12i (top), m12f (middle), and m12m (bottom).   
    The chemical difference distributions shown in each column are computed as using the method presented \S\ref{subsec: chemical difference metric}, but leverage different elements to compute the chemical difference metric. 
    The left panel is a reproduction of Fig.~\ref{fig: chemical difference metric }, which was generated using all relevant elements in our simulations (C, N, O, Mg, Si, S, Ca, Ne).
    The middle column shows the chemical difference distribution calculated using three elements (Mg, C, N).
    The right column shows the chemical difference distribution calculated using a difference subset of three elements (Mg, C, O).
    We note that decreasing the number of elements increases both the number of intra-cluster pairs and inter-cluster pairs that populate the smallest chemical difference metric bin.
    There is slightly more contamination from inter-cluster pairs when fewer elements are used to calculate the chemical difference metric.
    While broadly all of the distributions presented here look very similar, the maximum difference between the inter and intra-cluster global distributions are found when all elements are used to calculate the metric.}
    \label{fig: appendix metrics distribution }
\end{figure*}

We show histograms of the chemical difference metric in Fig.~\ref{fig: appendix metrics distribution } for pairs of stars within OCs indicated by blue histograms (intra-cluster pairs) and between OCs indicated by the dashed red histograms (inter-cluster pairs) for each simulation: m12i (top panel), m12f (middle panel), and m12m (bottom panel). 
These chemical difference distributions are computed similarly those presented in Fig.~\ref{fig: chemical difference metric }, which was generated by selecting OCs that fall within $\pm{1\sigma}$ from the overall mean distribution of [Fe/H] in each simulation and then applying the formula presented in \S\ref{subsec: chemical difference metric} to relevant pairs drawn for the selected OCs. 
The first column on the left shows the distribution of the chemical difference metric when all elements were used. 
The middle column and right columns are for the cases when only 3 elements are used: Mg, C, N and Mg, C, O.
We note that decreasing the number of elements increases both the number of intra-cluster pairs and
inter-cluster pairs that populate the smallest chemical difference metric bin. 
There is slightly more contamination from inter-cluster pairs when fewer elements are used to calculate the chemical difference metric.
While broadly all of the distributions presented here look very similar, the maximum difference between
the inter and intra-cluster global distributions are found when all elements are used to calculate the metric.

In Table \ref{Table: Appendix chemical difference metric using all and few elements stats}, we show the percentage of pairs that have a chemical difference metric smaller than the mean, median or a fixed value when different elements are used to calculate the chemical difference metric.
While there is a slight increase or decrease in the percentage of pairs falling below a chosen limit, the chemical difference overlap -- that is, the relative amount of inter-cluster contamination -- does not change significantly; there is always significant contamination at small values of the chemical difference metric.
In Table~\ref{Table:purity_percentage}, we present the purity (the percentage of intra-cluster pairs relative to the total pairs) of each sample for each metric. 
Each population is significantly  impure -- the contamination rate is typically 40\%-50\%. 
We conclude that no matter what metric we use or subset of elements we consider, we always have significant contamination of inter-cluster pairs.

We note that the mean and median of the chemical difference distribution decreases very slightly when few elements are considered, and it decreases significantly when mean and median of the metrics are computed when combining the intra and inter-clusters distributions into one global distribution; we present this for completeness in Table \ref{Table:chemical_difference_metric_statistics}.

\begin{table*}
    \centering
    \caption{Chemical difference metric analysis using all elements versus a subset of elements (Mg, C, N and Mg, C, O)}
    \begin{tabular}{|c|c|c|c|c|c|c|c|}
        \hline
        Simulation & Selection of & Intra Cluster & Intra Cluster & Intra Cluster & Inter Cluster & Inter Cluster & Inter Cluster \\
        & Chemical Difference & All Elements & Mg, C, N & Mg, C, O & All Elements & Mg, C, N & Mg, C, O \\
        & Metric & (\%) & (\%) & (\%) & (\%) & (\%) & (\%) \\
        \hline
        \multirow{6}{*}{m12i} & < Intra Mean         & 76.23 & 74.94 & 75.94 & 59.61 & 58.82 & 59.07 \\
			      & < Intra Median       & 50.00 & 50.00 & 50.00 & 31.60 & 33.23 & 32.16 \\
		 	      & < Intra+Inter Mean   & 81.68 & 81.64 & 82.21 & 66.63 & 66.39 & 66.85 \\
			      & < Intra+Inter Median & 59.54 & 58.55 & 59.24 & 40.46 & 41.45 & 40.76 \\
                              & < 0.00010 (Fixed)    & 70.47 & 81.66 & 86.79 & 52.36 & 66.41 & 72.92 \\
                              & < 0.00020 (Fixed)    & 86.18 & 92.95 & 95.57 & 73.20 & 82.13 & 87.21 \\
        \hline
        \multirow{6}{*}{m12f} & < Intra Mean         & 83.20 & 79.47 & 84.33 & 70.39 & 65.76 & 71.57 \\
			      & < Intra Median       & 49.97 & 49.97 & 49.97 & 27.77 & 28.66 & 28.25 \\
                              & < Intra+Inter Mean   & 86.94 & 83.98 & 88.07 & 79.11 & 74.24 & 80.71 \\
                              & < Intra+Inter Median & 62.08 & 60.42 & 61.36 & 37.92 & 39.58 & 38.64 \\
                              & < 0.00010 (Fixed)    & 47.54 & 61.66 & 72.28 & 25.34 & 40.77 & 50.92 \\
                              & < 0.00020 (Fixed)    & 65.93 & 76.74 & 85.40 & 43.03 & 61.01 & 73.47 \\
        \hline
        \multirow{6}{*}{m12m} & < Intra Mean         & 83.16 & 81.55 & 82.62 & 62.03 & 67.65 & 66.04 \\
			      & < Intra Median       & 50.00 & 50.00 & 50.00 & 17.65 & 19.79 & 16.58 \\
                              & < Intra+Inter Mean   & 85.83 & 83.16 & 84.76 & 72.19 & 72.19 & 72.46 \\
                              & < Intra+Inter Median & 67.11 & 63.64 & 65.24 & 32.89 & 36.36 & 34.76 \\
                              & < 0.00010 (Fixed)    & 70.59 & 79.14 & 82.62 & 38.77 & 61.50 & 67.11 \\
                              & < 0.00020 (Fixed)    & 83.69 & 86.63 & 89.04 & 64.71 & 79.68 & 83.16 \\
        \hline
    \end{tabular}
\label{Table: Appendix chemical difference metric using all and few elements stats}
\end{table*}

\begin{table*}
    \centering
    \caption{Purity percentage of intra-cluster pairs}
    \begin{tabular}{|c|c|c|c|c|}
        \hline
        Simulation & Statistic & All Elements & Mg, C, N & Mg, C, O  \\
        & & Purity ($\%$) & Purity ($\%$) & Purity ($\%$) \\
        \hline
        \multirow{6}{*}{m12i} & < Intra Mean         & 56.13 & 56.05 & 56.23 \\
                              & < Intra Median       & 61.29 & 60.09 & 60.91 \\
		 	                  & < Intra+Inter Mean   & 55.09 & 55.12 & 55.16 \\
		                      & < Intra+Inter Median & 59.54 & 58.55 & 59.24 \\
                              & < 0.00010 (Fixed)    & 57.36 & 55.14 & 54.36 \\
                              & < 0.00020 (Fixed)    & 54.08 & 53.12 & 53.21 \\
        \hline
        \multirow{6}{*}{m12f} & < Intra Mean         & 54.14 & 54.56 & 54.94 \\
                              & < Intra Median       & 64.69 & 64.52 & 64.15 \\
                              & < Intra+Inter Mean   & 54.13 & 53.34 & 53.57 \\
                              & < Intra+Inter Median & 62.08 & 60.42 & 61.36 \\
                              & < 0.00010 (Fixed)    & 47.54 & 61.66 & 72.28 \\
                              & < 0.00020 (Fixed)    & 65.93 & 76.74 & 85.40 \\
        \hline
        \multirow{6}{*}{m12m} & < Intra Mean         & 59.64 & 60.04 & 60.31 \\
                              & < Intra Median       & 55.23 & 54.91 & 54.84 \\
                              & < Intra+Inter Mean   & 56.03 & 55.47 & 55.23 \\
                              & < Intra+Inter Median & 67.11 & 63.64 & 65.24 \\
                              & < 0.00010 (Fixed)    & 57.34 & 55.65 & 55.12 \\
                              & < 0.00020 (Fixed)    & 54.39 & 53.76 & 53.92 \\
        \hline
    \end{tabular}
\label{Table:purity_percentage}
\end{table*}

\begin{table*}
    \centering
    \caption{Chemical difference metric statistics}
    \begin{tabular}{|c|c|c|c|c|}
        \hline
        Simulation & Statisic & All Elements & Mg, C, N & Mg, C, O \\
        \hline
        \multirow{4}{*}{m12i} &  Intra Mean         & 0.00007 & 0.00007 & 0.00006 \\
			      &  Intra Median       & 0.00003 & 0.00003 & 0.00002 \\
		 	      &  Intra+Inter Mean   & 0.00016 & 0.00010 & 0.00008 \\
			      &  Intra+Inter Median & 0.00007 & 0.00004 & 0.00003 \\

        \hline
        \multirow{4}{*}{m12f} & Intra Mean          & 0.00024 & 0.00024 & 0.00019 \\
			      & Intra Median        & 0.00006 & 0.00006 & 0.00004 \\
                              & Intra+Inter Mean    & 0.00065 & 0.00032 & 0.00026 \\
                              & Intra+Inter Median  & 0.00017 & 0.00009 & 0.00006 \\
                           
        \hline
        \multirow{4}{*}{m12m} & Intra Mean          & 0.00012 & 0.00012 & 0.00010 \\
			      & Intra Median        & 0.00002 & 0.00002 & 0.00002 \\
                              & Intra+Inter Mean    & 0.00024 & 0.00014 & 0.00012 \\
                              & Intra+Inter Median  & 0.00009 & 0.00005 & 0.00004 \\
                              
        \hline
    \end{tabular}
\label{Table:chemical_difference_metric_statistics}
\end{table*}

\label{lastpage}

\end{document}